\documentclass[journal]{IEEEtran}
\ifCLASSINFOpdf
\else
\fi
\usepackage{graphicx}
\usepackage{setspace}
\usepackage{amsmath}
\usepackage{amssymb} 
\usepackage{caption}
\usepackage{subcaption}
\usepackage{bm} 
\usepackage{cite}
\usepackage{float}
\usepackage[usenames,dvipsnames]{pstricks}
\usepackage{epsfig}
\usepackage{pst-grad} % For gradients
\usepackage{pst-plot} % For axes
\usepackage{tabularx}
\usepackage{flushend}
\usepackage[T1]{fontenc}
\newcommand{\lb} {\left}
\newcommand{\rb} {\right}
\newcommand{\nn} {\nonumber}
\usepackage{xcolor}
\usepackage{lipsum}

\allowdisplaybreaks

\allowdisplaybreaks
\flushend
\usepackage{caption}
\usepackage{subcaption}
\usepackage[utf8]{inputenc}

 \IEEEaftertitletext{\vspace{-2.5\baselineskip}}
\begin{document}

\onecolumn{\noindent This article is accepted for publication in IEEE Transactions on Vehicular Technology.
\\
\\
\noindent © 2022 IEEE. Personal use of this material is permitted. Permission from IEEE must be obtained for all other uses, in any current or future media, including reprinting/republishing this material for advertising or promotional purposes, creating new collective works, for resale or redistribution to servers or lists, or reuse of any copyrighted component of this work in other works.
}

\twocolumn{

\bstctlcite{IEEEexample:BSTcontrol}
\title{Ergodic Secrecy Rate of Optimal Source-Destination Pair Selection in Frequency-Selective Fading}
\author{
  \IEEEauthorblockN{Shashi Bhushan Kotwal\IEEEauthorrefmark{1}, Chinmoy Kundu\IEEEauthorrefmark{2}, 
  Sudhakar Modem\IEEEauthorrefmark{3},
  Ankit Dubey\IEEEauthorrefmark{4}, 
  and Mark F. Flanagan\IEEEauthorrefmark{5}
  }
  \IEEEauthorblockA{\IEEEauthorrefmark{1}\IEEEauthorrefmark{3}\IEEEauthorrefmark{4}Department of EE, Indian Institute of Technology Jammu, Jammu \& Kashmir, India }
  % \IEEEauthorblockA{\IEEEauthorrefmark{2}Department of Electrical Engineering, Indian Institute of Technology Jammu, Jagti, NH44, Jammu \& Kashmir 181221, India} 
  
  % \IEEEauthorblockA{\IEEEauthorrefmark{3}Department of Electrical Engineering, Indian Institute of Technology Jammu, Jagti, NH44, Jammu \& Kashmir 181221, India}
  % \IEEEauthorblockA{\IEEEauthorrefmark{4}School of Electronics, Electrical Engineering and Computer Science, Queen’s University Belfast, U.K.}
  \IEEEauthorblockA{\IEEEauthorrefmark{2}\IEEEauthorrefmark{5}School of Electrical and Electronic Engineering, University College Dublin, Belfield, Ireland}
  %option 1: % \textrm{\IEEEauthorrefmark{1}2019ree0010@iitjammu.ac.in}, {\IEEEauthorrefmark{2}chinmoy.kundu@ucd.ie},
  %  {\IEEEauthorrefmark{3} ankit.dubey@iitjammu.ac.in}
  %  {\IEEEauthorrefmark{4} jzhang22@qub.ac.uk}
  %  {\IEEEauthorrefmark{5} mark.flanagan@ucd.ie}
  %Option2
  \textrm{
  {{\IEEEauthorrefmark{1}sbkotwal@ieee.org},
  \{\IEEEauthorrefmark{3}sudhakar.modem ,\IEEEauthorrefmark{4}ankit.dubey\}
  @iitjammu.ac.in}, {\IEEEauthorrefmark{2}chinmoy.kundu@ucd.ie}, 
  {\IEEEauthorrefmark{5}mark.flanagan@ieee.org}}}
  \author{Shashi Bhushan Kotwal,~\IEEEmembership{Student Member,~IEEE,} Chinmoy Kundu,~\IEEEmembership{Member,~IEEE,} Sudhakar Modem,~\IEEEmembership{Member,~IEEE,} Ankit Dubey,~\IEEEmembership{Member,~IEEE,} and Mark F. Flanagan,~\IEEEmembership{Senior Member, IEEE}

\thanks{Shashi Bhushan Kotwal, Sudhakar Modem, and Ankit Dubey are with Indian Institute of Technology Jammu, India
% (email: \{2019ree0005@iitjammu.ac.in/sbkotwal@ieee.org\},\{sudhakar.modem, ankit.dubey\}@iitjammu.ac.in).}
(email: \{shashi.kotwal, sudhakar.modem, ankit.dubey\}@iitjammu.ac.in).}
\thanks{Chinmoy Kundu and Mark F. Flanagan are with University College Dublin, Ireland (email: chinmoy.kundu@ucd.ie and mark.flanagan@ieee.org).}
\thanks{This publication has emanated from research supported in part by Science Foundation Ireland (SFI) under Grant Number 17/US/3445 and is co-funded under the European Regional Development Fund under Grant Number 13/RC/2077. It is also co-funded by the Science and Engineering Research Board, India sponsored project ECR/2018/002795.
} 
% }% <-this % stops a space
% \thanks{Manuscript received April 19, 2005; revised December 27, 2012.}}
}

\maketitle

\thispagestyle{empty}
\begin{abstract}

Node selection is a simple technique to achieve diversity and thereby enhance the physical layer security in future wireless communication systems which require low complexity. High-speed data
transmission often encounters frequency selective fading. In this context, we evaluate the exact closed-form
expression for the ergodic secrecy rate (ESR) of the optimal source-destination pair selection
scheme with single-carrier cyclic-prefix modulation, where the destination and eavesdropper channels
both exhibit independent frequency selective fading with an arbitrary number of multipath components. A simplified analysis in the high-SNR scenario along with an asymptotic analysis is also provided. We also derive and compare the corresponding results for the sub-optimal source-destination pair
selection scheme. We show that our analysis produces the corresponding ESR results under narrowband
independent Nakagami-$m$ fading channels with any arbitrary integer parameter $m$. 
The effect of
transmitters, destination, and eavesdropping paths correlation on the ESR is also demonstrated. Our solution
approach is general and can be used to find the ESR of a wider variety of transmitter selection schemes.
\end{abstract}


\section{Introduction}\label{section_Introduction}
In future wireless networks, e.g. internet-of-things (IoT), where low-complexity nodes are preferable, physical layer security (PLS) may be a suitable choice for implementing data security with its relatively simple channel coding techniques, instead of implementing complex higher-layer encryption techniques \cite{Vincet_Poor_fund_of_IoT_ICC_2019}.
Improving PLS through the diversity gain achieved by an antenna, transmitter/source, relay, destination, and/or  source-destination pair selection technique is a relatively simple method suitable for implementation in low-complexity dense and heterogeneous future networks, and has been studied extensively in \cite{Mallik_TWC_14, Shilpa_ICCST19, Zhao_MIMO_outdated_CSI_TC18, Kim_2016_CPSC_Trans,Chinmoy_TWC_2015,Chinmoy_GC16, Chinmoy_TVT_2019, Shalini_GC20, Chinmoy_letters21, SBK_VTC21,Kim_PLS_dACDD_TC_2020,Kim_2015_CPSC,Kim_2018_CDD_JOUR,other_2017_CPSC}. The authors in \cite{Chinmoy_TWC_2015, Chinmoy_GC16,Mallik_TWC_14,Shilpa_ICCST19,Chinmoy_TVT_2019, Shalini_GC20, Chinmoy_letters21,Zhao_MIMO_outdated_CSI_TC18,Kim_2016_CPSC_Trans,Kim_PLS_dACDD_TC_2020,SBK_VTC21} considered only source or relay selection schemes. Other works considered joint source-relay pair\cite{other_2017_CPSC, Kim_2018_CDD_JOUR}  or relay-destination pair selection \cite{Kim_2015_CPSC} in a multiple relay/destination scenario. 

 % Other works considered joint relay-destination pair selection \cite{Kim_2015_CPSC} or source-relay pair\cite{Kim_2018_CDD_JOUR,other_2017_CPSC} selection in a multiple source/relay/destination scenario.
% The selection schemes can be broadly classified into two types, i.e. sub-optimal  selection (SS) and optimal  selection (OS) \cite{Chinmoy_TWC_2015}. 
In situations where only source to destination channel(s) state information (CSI) was available, the authors in \cite{ Mallik_TWC_14, Shilpa_ICCST19, Zhao_MIMO_outdated_CSI_TC18, Kim_2016_CPSC_Trans, other_2017_CPSC,Kim_2018_CDD_JOUR,Kim_2015_CPSC,Kim_PLS_dACDD_TC_2020} implemented sub-optimal node selection scheme. In addition to the source to destination CSI, the authors in \cite{Chinmoy_TWC_2015, Chinmoy_GC16,Chinmoy_TVT_2019,Shalini_GC20,Chinmoy_letters21,SBK_VTC21} assumed the availability of the source to eavesdropper CSI and implemented an optimal node selection scheme. The sub-optimal selection scheme is aimed at maximizing the source to destination channel rate,  whereas the optimal selection  scheme maximizes the overall secrecy rate of the system. Moreover, it can provide a theoretical upper bound on the performance of sub-optimal selection schemes.

As PLS exploits the random behavior of the channel, the nature of channel fading, i.e., narrowband flat fading or frequency selective fading, affects the performance of each selection scheme differently.
Most of the literature on node selection for security is studied in the context of a narrowband flat fading channel
\cite{Chinmoy_TWC_2015, Chinmoy_TVT_2019, Mallik_TWC_14, Shilpa_ICCST19, Chinmoy_GC16, Shalini_GC20, Chinmoy_letters21}. 
% Performance analysis under Rayleigh fading channel is prevalent, e.g., in \cite{Chinmoy_GC17, Chinmoy_TWC_2015, Chinmoy_TVT_2019, Chinmoy_GC16, Shalini_GC20, Chinmoy_letters21} and the references therein. 
% In these papers, the secrecy outage probability (SOP) is generally considered as principal performance metric.
The secrecy outage probability (SOP) is generally considered to be the main performance parameter in these papers.
% The SOP of various sub-optimal relay selection schemes in a cooperative relay system was presented in \cite{Chinmoy_GC17}. 
% For the cooperative system, the SOP analysis of optimal relay selection was carried out in \cite{Chinmoy_GC16, Chinmoy_TWC_2015}. 
The authors in 
% \cite{Mallik_TWC_14,KIM_PLS_Nakagami_TSP_2017,Zhao_MIMO_outdated_CSI_TC18,Shilpa_ICCST19}
\cite{Mallik_TWC_14,Shilpa_ICCST19} 
extended the SOP analysis to ergodic secrecy rate (ESR) analysis for a transmitter selection scheme in a Nakagami-$m$ fading environment. Although the authors presented a closed-form solution, the scheme is sub-optimal. Moreover, the authors in \cite{Shilpa_ICCST19} did not provide an exact ESR analysis but an upper bound; also, high-SNR and asymptotic analysis were not provided. SOP and ESR analysis of a sub-optimal transmitter antenna selection in a relayed network with maximal-ratio combining (MRC) at the relay, destination, and eavesdropper was performed in \cite{Zhao_MIMO_outdated_CSI_TC18} assuming outdated CSI. The authors showed that system secrecy performance degrades in the presence of outdated CSI. The analysis of optimal selection  is absent in \cite{Zhao_MIMO_outdated_CSI_TC18} and the methodology adopted for the ESR analysis in the sub-optimal selection  scheme is also different from our work. The authors showed that outdated CSI degrades the secrecy performance. The authors in \cite{Chinmoy_TVT_2019} analyzed the SOP for both the sub-optimal and optimal selection  schemes in a cognitive radio network. However, the exact ESR for the optimal selection  scheme was not presented. The optimal selection  scheme was also implemented in \cite{Shalini_GC20} on the same system as that of \cite{Chinmoy_TVT_2019} using machine learning techniques; however, no theoretical analysis was presented.
Only recently, the ESR of the optimal selection  scheme in the presence of multiple eavesdroppers was presented in \cite{Chinmoy_letters21} under Rayleigh fading channel conditions. However, the authors neither analyzed the performance of the source-destination pair selection schemes nor considered frequency selective fading channel conditions.

High-speed wireless data transmission often encounters multipath frequency selective fading in practice. In this scenario, single carrier-cyclic prefix (SC-CP) signaling has been proposed as a solution to overcome the effects of inter-symbol interference (ISI) \cite{OFDM_vs_SCCP}. Using SC-CP in a frequency selective fading channel, the authors in \cite{Kim_2015_CPSC,Kim_2016_CPSC_Trans,Kim_2018_CDD_JOUR,Kim_PLS_dACDD_TC_2020,other_2017_CPSC,SBK_VTC21} considered PLS enhancement through transmitter selection.
% In a cooperative communication with wireless backhaul involving a single relay and destination, and multiple eavesdroppers, the authors in \cite{Kim_2016_CPSC_Trans} derived the ESR (including performing an asymptotic analysis) for a sub-optimal selection scheme where the selection was performed on the basis of the transmitter to relay channel quality.
In a cooperative communication system with wireless backhaul involving a single relay and destination and multiple eavesdroppers, the authors in \cite{Kim_2016_CPSC_Trans} investigated the ESR based on the transmitter to relay channel quality for sub-optimal selection scheme [5].
In another cooperative communications scenario with multiple relays, multiple destinations, and multiple eavesdroppers, a two-stage joint relay-destination pair selection scheme was proposed in \cite{Kim_2015_CPSC}, where the relay was selected to minimize the  eavesdropping rate. Although ESR is evaluated, the selection scheme is  not aimed at maximizing the secrecy capacity of the system and is therefore sub-optimal.
 A system with wireless backhaul was considered in \cite{Kim_2018_CDD_JOUR}, where a sub-optimal transmitter-relay pair selection  scheme is implemented in the presence of multiple eavesdroppers; here selection was based on maximizing the source to destination channel rate. 
To degrade the eavesdropping channel with the help of a jammer in the presence of multiple eavesdroppers in a system including wireless backhaul, a two-stage transmitter-relay pair selection scheme was proposed in \cite{other_2017_CPSC}; this selection scheme was also sub-optimal. Only recently, the authors in \cite{SBK_VTC21} introduced the optimal selection  scheme for the context of a frequency selective fading channel. However, the authors consider the SOP as the performance metric, not the ESR.

The authors in \cite{PLS_TAS_Corr_2013, SOP_corr_Rayleigh_ICC_19} present the SOP analysis of transmitter antenna selection with correlated channels. The authors in \cite{PLS_TAS_Corr_2013} assume receive antenna correlation at the destination and  eavesdropper. It is shown that the receive antenna correlation is detrimental to the secrecy performance. The authors in \cite{SOP_corr_Rayleigh_ICC_19} showed that the correlation between destination and eavesdropper channels benefits the SOP performance at high SNR. Neither of these papers considers frequency selective fading or ESR performance. If the scattering clusters causing multipath components are at a large distance from the transmitter, but in a narrow angular range, path correlation can be separated from the transmitter/transmit antenna correlation \cite{Space_tap_Corr_TWC_2007}. The authors in  \cite{Space_tap_Corr_TWC_2007} showed that in frequency selective channels, the transmitter and receiver antenna correlation along with path correlation have a detrimental effect on the channel capacity. However, the effect of path correlation on the secrecy rate in frequency selective fading channel is unexplored. 

In next-generation dense heterogeneous networks, multiple source radio nodes need to serve multiple receive radio nodes simultaneously. The source nodes may be acting as relaying nodes for the next level of receiving nodes in a multi-hop communication setup. In this scenario, choosing the best forwarding path among all possible available paths in order to maximize secrecy is required \cite{Kim_2015_CPSC,other_2017_CPSC,Kim_2018_CDD_JOUR}. Another practical scenario in which adaptive source-destination pair scheduling is implemented maximizing the secrecy of the system is described in  \cite{No_PLS_Sched_VTC_Sep_2004,PLS_Sched_TC_13,PLS_Sched_cellular_IEEE_NET_15}. However, in these scenarios, to the best of the author's knowledge, the ESR performance of the optimal source-destination pair selection (OS) scheme in frequency selective fading conditions is unavailable. Moreover, the effect of transmitter and path correlation on the ESR performance is not available in the literature. Moreover, the effect of transmitter and path correlation on the ESR performance is not available in the literature. This motivates us to consider the OS scheme and sub-optimal source-destination pair selection (SS) scheme in frequency selective channels.

The system models and the sub-optimal selection schemes in \cite{Kim_2015_CPSC,Kim_2016_CPSC_Trans,Kim_2018_CDD_JOUR,other_2017_CPSC} are different from that considered in this paper.
The two-stage transmitter-relay pair selection scheme in  \cite{Kim_2015_CPSC} does not consider the maximum source to destination channel rate, and a high-SNR analysis is also not presented. The work in \cite{Kim_2016_CPSC_Trans} involves relaying, and the selection is based on the transmitter to relay channel only. Hence, the selection scheme is different from our work, as it does not consider the maximum source to destination channel rate (also note that our results cannot be directly derived from \cite{Kim_2016_CPSC_Trans}). The system in \cite{Kim_2018_CDD_JOUR} also involves relaying, and as a result, the system is different from that considered in this paper; the asymptotic analysis of the ESR is also absent. In another two-stage transmitter-relay pair selection scheme (not-optimal) in \cite{other_2017_CPSC}, the transmitter in the first stage is selected based on source to relay channel rate instead of the source to destination channel rate as in our work. Also, the asymptotic analysis of the ESR is presented in integral form instead of closed form.  Although the authors in \cite{SBK_VTC21} analyze a transmitter selection scheme in frequency selective channel with SC-CP for both optimal and sub-optimal selection schemes, in contrast to our paper, it does not consider either source-destination pair selection or ESR analysis.
\begin{itemize}

    \item For the first time in the literature, we obtain the exact closed-form ESR for the OS scheme in a frequency selective fading channel with SC-CP modulation in a system consisting of multiple transmitters, multiple destinations, and an eavesdropper. The generalized expressions apply to an arbitrary number of transmitters and destinations, and an arbitrary number of multipath components in the destination and eavesdropper channels.

    \item We also derive the exact closed-form ESR of the SS scheme and compare it with that of the OS scheme.

    \item To reduce the computational complexity of the ESR and to find better insights, we derive a high-SNR approximation and asymptotic analysis of the ESR for both OS and SS schemes. We show that it is better for the system secrecy to increase the number of transmitters than the number of destinations in the case of the OS scheme. In contrast, in the case of the SS scheme, the number of transmitters and destinations have an equal effect on the ESR.

    \item To derive the ESR, we adopt a new approach with respect to the existing literature, which makes use of the cumulative distribution function (CDF) of the ratio of the destination to eavesdropper channel SNR. The method is general and can also be used to derive the ESR of a wide variety of node selection schemes, such as those presented in \cite{Chinmoy_GC17,Kim_2015_CPSC,Kim_2016_CPSC_Trans,Kim_2018_CDD_JOUR,other_2017_CPSC}. 

  \item We demonstrate the effect of the transmitter, destination, and eavesdropper path correlation on the ESR of the OS and SS schemes numerically. We conclude that correlation among transmitters is detrimental to the ESR performance, while path correlation improves it.
\end{itemize}

The rest of the paper is organized as follows. The system model is described in Section \ref{sec_System_Model}. The ESR of the OS scheme, together with its high-SNR and asymptotic analysis, are provided in Section \ref{sec_OS}, while the corresponding analyses for the SS  scheme are provided in Section \ref{sec_SS}. The transmitter and path correlation models are presented in Section \ref{sec_Spatial_Correlation}. Results and discussions are presented in Section \ref{sec_Results} and conclusions are drawn in Section \ref{sec_Conclusion}.

\textit{Notation:}  $||\mathbf{h}||$ denotes the Euclidean norm of a vector $\mathbf{h}$,  $|\mathcal{A}|$ denotes size of set $\mathcal{A}$, $\mathbb{P[\cdot]}$ denotes the probability of an event, $\otimes$ and $\mathbb{E}\left[\cdot \right]$ denote the Kronecker product and expectation operator respectively. $\text{vec}\{\cdot\}$, T and H denote vectorization, transpose, and Hermitian operator, respectively. $\Gamma(x)=\int_0^\infty t^{z-1}\exp(-t)dt$ denotes the complete Gamma function, while $\gamma(z,x)=\int_0^x t^{z-1}\exp(-t)dt$  and   $\Gamma(z,x)=\int_x^\infty t^{z-1}\exp(-t)dt$ denote the lower and upper incomplete Gamma function respectively. 
% and $\binom{m}{n}$ denotes the binomial coefficient. 
The probability density function (PDF) and the cumulative distribution function (CDF) of a random variable $X$ are denoted by $f_{X}(\cdot)$ and $F_{X}(\cdot)$, respectively.

\section{System Model}\label{sec_System_Model}
\begin{figure}\label{system_model}
 \centering
\includegraphics[width=10cm,height=5cm,keepaspectratio]{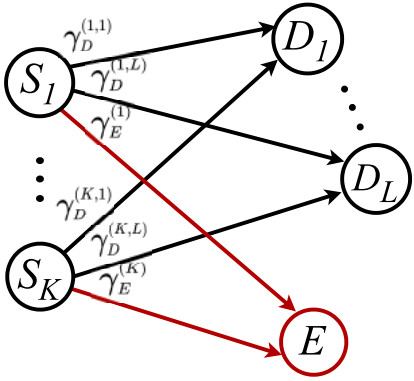} 
 \vspace{-0.2cm}
 \caption{System model consisting of $K$ transmitters, $L$ legitimate destinations, and an eavesdropper.}
 \vspace{-0.5cm}
 \end{figure}
We assume that $K$ transmitters $S_k$, where  $k\in\mathcal{S}=\{1 ,\ldots, K\}$, are transmitting data to $L$ legitimate destinations $D_l$, where $l\in\mathcal{D}=\{1 ,\ldots, L\}$, through wireless channels in the presence of a passive eavesdropper $E$. 
% We consider a generalized model with multiple destinations to improve the secrecy performance as compared to the case of  single destination.
% We assume a passive eavesdropper in the system which is able to overhear the communication meant for the destinations.
We consider SC-CP modulation at each transmitter, assuming a frequency selective fading channel between each pair of nodes.
% to overcome inter symbol interference (ISI). 
Due to the multipath propagation, the destination channel $S_k$-$D_l$ for each $k \in \mathcal{S}$ and $l \in \mathcal{D}$, and the eavesdropping channel $S_k$-$E$ for each $k \in \mathcal{S}$,  contain $M_{D}$ and $M_{E}$ paths, respectively.  The destination channel $\mathbf{h}_{D}^{(k,l)}\in\mathbb{C}^{1 \times M_{D}}$ between $S_k$-$D_l$ is modelled as a circularly symmetric complex Gaussian random vector of length $M_{D}$ with zero mean and identity covariance matrix. Similarly, the eavesdropping channel $\mathbf{h}_{E}^{(k)}\in\mathbb{C}^{1 \times M_{E}}$ between $S_k$-$E$ is also modelled as a circularly symmetric complex Gaussian random vector of length $M_{E}$ with zero mean and identity covariance matrix. Therefore, the channel coefficient $h_{D}^{(k,l)}(i)$ for each path  $i \in\{1,2,\dots,M_{D}\}$, as well as $h_{E}^{(k)}(i)$ for each path $i \in\{1,2,\dots,M_{E}\}$, is an independent circularly symmetric complex Gaussian random variable with zero mean and unit variance, and its magnitude follows a Rayleigh distribution. 
We also assume that the channels  $\mathbf{h}_{D}^{(k,l)}$, for all $k$ and $l$ are  are independent and identically distributed (i.i.d.). Similarly channels  $\mathbf{h}_{E}^{(k)}$, for all $k$  are also i.i.d. However, the parameters of the distribution of the channels $\mathbf{h}_{D}^{(k,l)}$ and $\mathbf{h}_{E}^{(k)}$ are assumed to be different, in general.
Owing to SC-CP modulation, the SNRs for each $S_k$-$D_l$ and $S_k$-$E$ channel are written as \cite{Kim_2015_CPSC}
\begin{align}\label{gamma_Dk}
\gamma_{D}^{(k,l)} = \frac{a_{D} P_T }{{\sigma^{2}_{D}}} ||\mathbf{h}^{(k,l)}_{D}||^2
\end{align}
and
\begin{align}\label{gamma_Ek}
\gamma_{E}^{(k)} = \frac{ a_{E} P_T }{{\sigma^{2}_{E}}} ||\mathbf{h}^{(k)}_{E}||^2, 
\end{align}
respectively, where $P_T$ is the transmitted power, $a_{D}$ and $a_{E}$ are the path loss factors which are inversely proportional to the $S_k$-$D_l$ and $S_k$-$E$ distance respectively for each $k$ and $l$, $\sigma^{2}_{D}$ and $\sigma^{2}_{E}$  are the average noise powers at $D_l$ and $E$, respectively. The PDF of $\gamma_{D}^{(k,l)}$ follows a Gamma distribution as \cite{Kim_2015_CPSC}
 \begin{align}\label{pdf_SCCP}
 f_{\gamma_{D}^{(k,l)}}(x) = \frac{{x}^{M_{D}-1} \exp({-\frac{x}{a_{D} P_T/{\sigma_{D}^2}}})}{\left({ a_D P_T/{\sigma_D^2}}\right)^{M_{D}}\Gamma(M_{D})} =\frac{x^{M_{D}-1}\exp{(-\frac{x}{\lambda_{D}})}}{\lambda_{D}^{M_{D}}\Gamma(M_{D})},
\end{align}
where  $\lambda_{D}=a_DP_T/\sigma_{D}^2$. As observed in \cite{SBK_VTC21}, the underlying frequency selective channel upon SC-CP modulation behaves as a narrowband Nakagami-$m$ fading channel with integer shape parameter $m=M_{D}$, and scale parameter representing average SNR     $\mathbb{E}[\gamma_{D}^{(k,l)}]=M_{D}\lambda_{D}$, where the average SNR per multipath component is $\lambda_{D}$. As the channel behaves as a Nakagami channel, the ESR performance derived for the OS and SS schemes in this paper is also applicable for the Nakagami fading channel\footnote{
% exactly, by tuning the shape and scale parameters, Nakagami-$m$ fading can represent the empirical channel data more accurately \cite{Nakagami_rayleigh_letters_2003}. 
The Rayleigh faded channel with $L$-branch maximal ratio combining (MRC) results in Nakagami fading with shape  parameter $m = L$. Further, the MRC combining $L$-branch Nakagami-$m$ faded signals results in another Nakagami faded signal with shape parameter $mL$ \cite{relation_Rayleigh_Nakagami_MRC}. Therefore, the analysis provided in this paper can be thought of as a generalized solution that can also lead to solutions for the Nakagami fading context} in  similar system assumptions  (note that this analysis is not currently available in the literature). As Rayleigh fading is a particular case of Nakagami-$m$ fading when shape parameter $m=1$, the results derived in this paper also apply to the Rayleigh fading channel.

The CDF of $\gamma_{D}^{(k,l)}$ is in the form of an incomplete Gamma function.  

\begin{align}\label{CDF_Dest}
F_{\gamma_{D}^{(k,l)}}(x) &=\int_0^x\frac{y^{M_{D}-1}\exp{(-\frac{y}{\lambda_{D}})}}{\lambda_{D}^{M_{D}}\Gamma(M_{D})}dy
% \nn\\
% &
=\frac{\gamma\Big(M_D,\frac{x}{\lambda_{D}}\Big)}{\Gamma(M_D)}.
\end{align}
%  where $\gamma(z,x)=\int_0^x y^{z-1} \exp({-y})dy$.
As $M_{D}$ is an integer, the CDF of $\gamma_{D}^{(k,l)}$ can be written as \cite[eq. (8.352.6)]{ryzhik_2007}
\begin{align}\label{CDF_S_D}
F_{\gamma_{D}^{(k,l)}}(x) &=1-\exp{\Big(-\frac{x}{\lambda_{D}}\Big)}\sum_{m=0}^{M_{D}-1}\frac{1}{m!}\Big(\frac{x}{\lambda_{D}}\Big)^{m}.
\end{align}
The PDF and CDF of $\gamma_{E}^{(k)}$ can be expressed by replacing  $D$ with $E$ and $(k,l)$ with $(k)$ in \eqref{pdf_SCCP} and \eqref{CDF_S_D}, respectively.
We assume that the destination and eavesdropper channels are non-identical, i.e., $\lambda_D$ and $M_D$ are not necessarily equal to $\lambda_E$ and $M_E$, respectively.
The secrecy rate achievable for the $k^{\textrm{th}}$ transmitter-$l^{\textrm{th}}$ destination pair in bits per channel use (bpcu) is defined as \cite{Wyner_1975}
\begin{align}
    C_{S}^{(k,l)} = \left[\log_2\big(\Gamma_{S}^{(k,l)}\big)\right]^+,
\end{align}
where 
\begin{align}
\label{eq_Gamma_sk}
\Gamma_{S}^{(k,l)}=\frac{1+\gamma_{D}^{(k,l)}}{1+\gamma_{E}^{(k)}}.
\end{align}
The corresponding ESR  
% to the $k$-th transmitter 
is then evaluated as \cite{Chinmoy_letters21}
 \begin{align}\label{eq_ESR_basic_eqn}
     C_{\mathrm{erg}}^{(k,l)} &= \mathbb{E}\big[C_{S}^{(k,l)}\big] = \frac{1}{\text{ln}(2)}\int_1^\infty \ln{(x)}f_{\Gamma_{S}^{(k,l)}}(x) dx. 
    %   C_{erg} &= \mathbb{E}\left[C_{S}\right] = \frac{1}{\text{ln}(2)}\int_1^\infty \ln{(x)}f_{\Gamma_{S}}(x) dx.
 \end{align}
 By changing the order of integration and after performing some mathematical manipulations, $C_{erg}^{(k)}$ is  reformulated to utilize the CDF of $\Gamma_{S}^{(k,l)}$ as
\begin{align}\label{eq_ESR_CDF_eqn}
 C_{\mathrm{erg}}^{(k,l)} &= \frac{1}{\text{ln}(2)}\int_1^\infty \frac{1-F_{\Gamma_{S}^{(k,l)}}(x)}{x} dx. 
\end{align}
The ESR evaluation using \eqref{eq_ESR_CDF_eqn} is simpler than using \eqref{eq_ESR_basic_eqn}, as \eqref{eq_ESR_CDF_eqn} does not require derivation of the PDF $ f_{\Gamma_{S}^{(k,l)}}(x)$ in (\ref{eq_ESR_basic_eqn}). Moreover, the analytical methodology proposed using the CDF of $\Gamma_{S}^{(k,l)}$ as in \eqref{eq_ESR_CDF_eqn} makes the ESR derivation of the OS scheme possible. In contrast, existing methodologies used in \cite{Mallik_TWC_14,Shilpa_ICCST19, Kim_2015_CPSC,Kim_2016_CPSC_Trans,Kim_2018_CDD_JOUR,other_2017_CPSC},  which are sub-optimal, consider distributions of the destination channel and eavesdropper channel separately, and hence, cannot capture the distribution of the maximal ratio of the destination to eavesdropper channel SNRs as we will do in Section \ref{sec_CDF_OS}. 
The method of formulating the ESR as in (\ref{eq_ESR_CDF_eqn}) with the help of the CDF of $\Gamma_{S}^{(k,l)}$ has been explored only in
% \cite{Mallik_TWC_14,Shilpa_ICCST19} or \cite{Kim_2015_CPSC,Kim_2016_CPSC_Trans,Kim_2018_CDD_JOUR,other_2017_CPSC}
% except in
\cite{Chinmoy_letters21}; however, \cite{Chinmoy_letters21} considers a different system model and a simple narrowband Rayleigh fading channel. In this paper, we face additional challenges due to the arbitrary number of destination nodes and the arbitrary number of multipath components in the destination and eavesdropper channels. In contrast, there is only one destination and one channel path in \cite{Chinmoy_letters21}. The arbitrary number of destination and multipath components makes it difficult to transform $F_{\Gamma_{S}^{(k,l)}}(x)$ into a suitable form to evaluate the integral in (\ref{eq_ESR_CDF_eqn}) and thus obtain a closed-form ESR solution.

 Our method is also general and can be used to evaluate the ESR of more general selection schemes when the distribution of (\ref{eq_Gamma_sk}) and the integration in (\ref{eq_ESR_basic_eqn}) are obtained in closed form. Hence, we also derive the ESR of the SS scheme, which maximizes the destination channel rate and does not consider the eavesdropping channel rate using (\ref{eq_ESR_CDF_eqn}), and compare it with that of the OS scheme. 

On a side note, the CDF of $\Gamma_{S}^{(k,l)}$ can also be used to evaluate various other performance metrics such as SOP, probability of non-zero secrecy rate (PNZ), secrecy throughput \cite{Effective_Secrecy_throughput_ICC_2014}, etc. of various node selection schemes including minimum and active eavesdropping \cite{Chinmoy_GC17}, SS (system model considered in \cite{Kim_2015_CPSC,Kim_2016_CPSC_Trans,Kim_2018_CDD_JOUR,other_2017_CPSC} and in our work), and OS.

\section{ESR of Optimal Source-Destination Pair  Selection}\label{sec_OS}
In this section, we present the ESR analysis of the OS scheme, its high-SNR analysis, and its asymptotic limit.
The OS scheme corresponds to selecting the source-destination pair for which the instantaneous secrecy rate of the system is maximized. Hence, from (\ref{eq_Gamma_sk}), the
OS scheme satisfies
\begin{align}\label{eq_Gamma_s}
\Gamma_{S}&=\max_{\forall k, \forall l}\big\{\Gamma_{S}^{(k,l)}\big\}=\max_{\forall k, \forall l}\Bigg\{\frac{1+\gamma_{D}^{(k,l)}}{1+\gamma_{E}^{(k)}}\Bigg\}\nn\\
&=\max_{\forall k}\Bigg\{\frac{1+\max\limits_{\forall l}\left\{\gamma_{D}^{(k,l)}\right\}}{1+\gamma_{E}^{(k)}}\Bigg\}
.
%\nn\\
\end{align}

The ESR for the OS scheme can be evaluated via  \eqref{eq_ESR_CDF_eqn} with the help of the CDF of $\Gamma_{S}$. Thus, we proceed to find the CDF of $\Gamma_{S}$ in the following subsection.
% \vspace{-0.3cm}
\subsection{Determining the CDF of  $\Gamma_{S}$ }\label{sec_CDF_OS}
Our intention in this section is to find the CDF of  $\Gamma_{S}$ in a suitable form that is tractable and easily integrable in \eqref{eq_ESR_CDF_eqn}.  By considering $\Gamma_{S}^{(k,l)}$ as i.i.d. random variables for all $k\in \mathcal{S}$ and $l\in \mathcal{D}$ in \eqref{eq_Gamma_sk}, the CDF of $\Gamma_{S}$ is determined following the basic definition of the CDF as 
\begin{align}\label{eq_CDF_OS_Psi_beginning_other}
  & F_{\Gamma_{S}}(x) =\mathbb P\lb[\Gamma_{S}\le x\rb]\nn\\
  &= \lb[\int_0^\infty F_{\max\limits_{\forall l}\{ \gamma_{D}^{(k,l)}\}}\left(x\left(1+y\right)-1\right)f_{\gamma_{E}^{(k)}}(y) dy\rb]^{K}\\
    & = \lb[\int_0^\infty \left(F_{ \gamma_{D}^{(k,l)}}\left(x\left(1+y\right)-1\right)\right)^L f_{\gamma_{E}^{(k)}}(y) dy\rb]^{K}\nn\\
    \label{eq_CDF_OS_Psi_beginning_old}
    &=\Bigg[1-\frac{1}{{\lambda_E}^{M_E}\Gamma(M_E)}\sum_{l=1}^{L}(-1)^{l+1}\binom{L}{l}\exp\left(-\frac{l(x-1)}{\lambda_D}\right) \nn\\
   & \times \int_0^\infty\underbrace{\Bigg(\sum_{m=0}^{M_D-1}\sum_{n=0}^{m}\sum_{u=0}^{m-n}(-1)^u\binom{m}{n}\binom{m-n}{u}\frac{x^{m-u}y^{n}}{m!{\lambda_D}^{m}}\Bigg)^l}_{\text{To be expanded using \eqref{eq_cross_mul_3_sums}}}\nn\\
   &\times y^{M_E-1}\exp\left(-\left(\frac{lx}{\lambda_D}+\frac{1}{\lambda_E}\right)y\right) dy\Bigg]^K\\
   \label{eq_CDF_OS_Psi_beginning}
   &= \left[1-\Psi(x)\right]^K,
% &= 1-\sum_{k=1}^{K}(-1)^{k+1}\binom{K}{k}\Psi^k(x),
\end{align}
where
\begin{align}\label{eq_Psi_OS_before_expansion}
    &\Psi(x)=\sum_{l=1}^{L}\sum_{\mathbf{m} \in \mathcal{M}^{(l)}} \sum_{\mathbf{n} \in \mathcal{N}^{(l)}(\mathbf{m})}\sum_{\mathbf{u} \in \mathcal{U}^{(l)}(\mathbf{m},\mathbf{n})}(-1)^{l+\widehat{u}_{p}^{(l)}+1}\binom{L}{l}\nn\\
    & \times \frac{{\lambda_D}^{M_E-(\widehat{m}^{(l)}-\widehat{u}^{(l)})}\Gamma\left(M_E+\widehat{n}^{(l)}\right)\Big(\prod\limits_{p=1}^{l}\binom{m_{p}}{n_p}\binom{m_{p}-n_p}{u_p}\Big)}{{\lambda_E}^{M_E}l^{M_E+\widehat{n}^{(l)}}\Big(\prod\limits_{p=1}^{l}{m_{p}}!\Big)\Gamma(M_E)}\nn\\
    & \times \frac{x^{\widehat{m}^{(l)}-\widehat{u}^{(l)}}\exp({-\frac{l\left(x-1\right)}{\lambda_{D}}})}{\big(x+\frac{\lambda_{D}}{l\lambda_{E}}\big)^{M_E+\widehat{n}^{(l)}}} , 
\end{align}
and using definitions 
\begin{align}
 \widehat{m}^{(l)}\triangleq\sum_{p=1}^{l}{m_{p}};~~~~ \widehat{n}^{(l)}\triangleq\sum_{p=1}^{l}{n_p};~~~~ \widehat{u}^{(l)}\triangleq\sum_{p=1}^{l}{u_p}.
 \end{align}
The expression in \eqref{eq_CDF_OS_Psi_beginning} and subsequently \eqref{eq_Psi_OS_before_expansion} is obtained using first the result of the form
\begin{align}\label{eq_cross_mul_3_sums}
&\Big(\sum_{i=0}^{\mu}\sum_{j=0}^i \sum_{v=0}^{i-j} f(i,j,v)\Big)^{\zeta}\nn\\
 &=\sum_{m_{1}=0}^{\mu}\sum_{n_1=0}^{m_{1}}\sum_{u_1=0}^{m_{1}-n_1}\ldots \sum_{m_{\zeta}=0}^{\mu}\sum_{n_{\zeta}=0}^{m_{\zeta}}\sum_{u_{\zeta}=0}^{m_{\zeta}-n_\zeta}\prod_{p=1}^{{\zeta}}f(m_{p},n_p,u_p)\nonumber\\
     &=\sum_{\mathbf{m} \in \mathcal{M}^{(\zeta)}} \sum_{\mathbf{n} \in \mathcal{N}^{(\zeta)}(\mathbf{m})}\sum_{\mathbf{u} \in \mathcal{U}^{(\zeta)}(\mathbf{m},\mathbf{n})}\prod_{p=1}^{{\zeta}}f(m_{p},n_p,u_p), 
\end{align} 
and then with the help of an integral solution of the form \cite[eq. (3.351.3)]{ryzhik_2007} in (\ref{eq_CDF_OS_Psi_beginning_old}). In (\ref{eq_cross_mul_3_sums}),  $f(i,j,v)$ is an arbitrary function of integers $i$, $j$ and $v$, $\mathcal{M}^{(\zeta)}$ is the set of integer vectors $[m_{1} ,\ldots,  m_{\zeta}]$ containing $\zeta$ elements  such that $m_{p}\in\{0 ,\ldots, \mu\}$ for each $p\in\{1 ,\ldots, \zeta\}$, $\mathcal{N}^{(\zeta)}(\mathbf{m})$ is the set of index vectors $[n_1 ,\ldots,  n_\zeta]$ containing $\zeta$ elements such that $n_p\in\{0 ,\ldots, m_{p}\}$, and  $\mathcal{U}^{(\zeta)}(\mathbf{m},\mathbf{u})$ is the set of integer vectors $[u_1 ,\ldots,  u_\zeta]$ containing $\zeta$ elements such that $u_p\in\{0 ,\ldots, \small(m_{p}-n_p\small)\}$.
% \textcolor{red}{Also we use $\mu=M_D-1$ and $\zeta =l$ in \eqref{eq_cross_mul_3_sums}(why are you writing it here in general equation? write where necessary in 12)}.  
Through binomial expansion, \eqref{eq_CDF_OS_Psi_beginning} is expressed as
\begin{align}\label{eq_CDF_OS_Psi}
   F_{\Gamma_{S}}(x) &= 1-\sum_{k=1}^{K}(-1)^{k+1}\binom{K}{k}\Psi^k(x).
\end{align}
Next, we transform $\Psi^k(x)$ in (\ref{eq_Psi_OS_expanded}) with the help of $\Psi(x)$ in (\ref{eq_Psi_OS_before_expansion}) using the following result similar to  \eqref{eq_cross_mul_3_sums}:  
\begin{align}\label{eq_cross_mul_4_sums}
&\Bigg(\sum_{l=1}^{L}\sum_{\mathbf{m} \in \mathcal{M}^{(l)}}  \sum_{\mathbf{n} \in \mathcal{N}^{(l)}(\mathbf{m})}\sum_{\mathbf{u} \in \mathcal{U}^{(l)}(\mathbf{m},\mathbf{n})}f(l,m_{p},n_p,u_p)\Bigg)^k \nn\\
&=
% \sum_{\mathbf{l},\mathbf{m},\mathbf{n},\mathbf{u}} 
\sum_{\mathbf{l} \in \mathcal{L}^{(k)}}\sum_{\mathbf{m}\in\mathcal{M}^{(k,\mathbf{l})}} \sum_{\mathbf{n}\in\mathcal{N}^{(k,\mathbf{l})}(\mathbf{m})}\sum_{\mathbf{u}\in\mathcal{U}^{(k,\mathbf{l})}(\mathbf{m},\mathbf{n})}\nn\\
&\left(\prod_{q=1}^{k}f(l_q,m_{q,p},n_{q,p},u_{q,p})\right), 
\end{align}
where
$\mathcal{L}^{(k)}$ is the set of index vectors $[l_1 ,\ldots,  l_k]$ containing $k$ elements such that $l_q\in\{1 ,\ldots, L\}$ for each $q\in\{1 ,\ldots, k\}$, $\mathcal{M}^{(k,\mathbf{l})}$ is a set of index vectors $[m_{1,1} ,\ldots,  m_{k,l_i}]$ containing $k \times l_i$ elements such that $m_{q,p} \in \{0 ,\ldots, M_D-1\}$ for each $q\in\{1 ,\ldots, k\}$ and $p\in\{1 ,\ldots, l_i\}$ for each $i\in\{1 ,\ldots, k\}$, $\mathcal{N}^{(k,\mathbf{l})}(\mathbf{m})$ is the set of index vectors $[n_{1,1} ,\ldots,  n_{k,l_i}]$ containing $k \times l_i$ elements such that $n_{q,p}\in\{0 ,\ldots, m_{q,p}\}$ for each $q\in\{1 ,\ldots, k\}$ and $p\in\{1 ,\ldots, l_i\}$ for each $i\in\{1 ,\ldots, k\}$,  $\mathcal{U}^{(k,\mathbf{l})}(\mathbf{m,n})$ is the set of index vectors $[u_{1,1} ,\ldots,  u_{k,l_i}]$ containing $k \times l_i$ elements such that $u_{q,p}\in\{0 ,\ldots, m_{q,p}-n_{q,p}\}$ for each  $q\in\{1 ,\ldots, k\}$ and $p\in\{1 ,\ldots, l_i\}$ for each $i\in\{1 ,\ldots, k\}$ and $f(l,m_{p},n_p,u_p)$ is an arbitrary function of integers $l$, $m_p$, $n_p$ and $u_p$.

Denoting by $\mathcal{X}$ the set of vector tuples $(\mathbf{l},\mathbf{m},\mathbf{n},\mathbf{u})$ described above, $\Psi^{k}(x)$ in \eqref{eq_CDF_OS_Psi} can be rewritten using \eqref{eq_cross_mul_4_sums} in a form suitable for finding the ESR, i.e.,
% We obtain $\Psi^k(x)$ in \eqref{eq_CDF_OS_Psi} into a form suitable for finding ESR easily as
\begin{align}\label{eq_Psi_OS_expanded}    
\Psi^k(x)&=\sum_{\lb(\mathbf{l},\mathbf{m},\mathbf{n},\mathbf{u}\rb)\in\mathcal{X}}\Upsilon\frac{x^{\widetilde{m}^{(k)}-\widetilde{u}^{(k)}}\exp({-\frac{\widetilde{l}^{(k)}}{\lambda_{D}}x})}{\prod\limits_{q=1}^{k}\big(x+\frac{\lambda_{D}}{l_q\lambda_{E}}\big)^{M_E+\widehat{n}_{q}^{(l_q)}}},
% \nn\\
% \Psi^k(x)&=\sum_{\mathbf{l},\mathbf{m},\mathbf{n},\mathbf{u}}\Upsilon\frac{x^{\widetilde{m}^{(k)}-\widetilde{u}^{(k)}}\exp({-\frac{\widetilde{l}^{(k)}}{\lambda_{D}}x})}{\prod\limits_{q=1}^{k}\big(x+\frac{\lambda_{D}}{l_q\lambda_{E}}\big)^{M_E+\widehat{n}_{q}^{(l_q)}}},
\end{align}
where
\begin{align}
% \label{eq_tilde_sum}
% \sum_{\lb(\mathbf{l},\mathbf{m},\mathbf{n},\mathbf{u}\rb)\in\mathcal{X}}&\triangleq \sum_{\mathbf{l} \in \mathcal{L}^{(k)}}\sum_{\mathbf{m}\in\mathcal{M}^{(k,\mathbf{l})}} \sum_{\mathbf{n}\in\mathcal{N}^{(k,\mathbf{l})}(\mathbf{m})}\sum_{\mathbf{u}\in\mathcal{U}^{(k,\mathbf{l})}(\mathbf{m},\mathbf{n})}\\
\label{eq_prod_short}
\Upsilon&\triangleq\prod\limits_{q=1}^{k}(-1)^{l_q+\widehat{u}_{q}^{(l_q)}+1}\binom{L}{l_q}\nn\\ 
&\times \frac{{\lambda_D}^{M_E-(\widehat{m}_{q}^{(l_q)}-\widehat{n}_{q}^{(l_q)})}\Gamma\big(M_E+\widehat{n}_{q}^{(l_q)}\big)}{{\lambda_E}^{M_E}{l_q}^{M_E+\widehat{n}_{q}^{(l_q)}}}\nn\\
& \times\frac{\big(\prod\limits_{p=1}^{l_q}\binom{m_{{q,p}}}{n_{q,p}}\binom{m_{{q,p}}-n_{q,p}}{u_{q,p}}\big)\exp\big({\frac{{l}_q}{\lambda_{D}}}\big)}{\big(\prod\limits_{p=1}^{l_q}m_{{q,p}}!\big)\Gamma(M_E)},
\end{align}
% with $\mathcal{X}$ defined as a set of integer vectors $[\mathbf{l},\mathbf{m},\mathbf{n},\mathbf{u}]$ 
and  
\begin{align}\label{eq_definition_OS_2}
\widetilde{l}^{(k)}&\triangleq\sum_{q=1}^{k}{l_q};~~~\widetilde{m}^{(k)}\triangleq\sum_{q=1}^{k}\sum_{p=1}^{l_q}{m_{{q,p}}};\widetilde{u}^{(k)}\triangleq\sum_{q=1}^{k}\sum_{p=1}^{l_q}{u_{q,p}}\nn\\
\widehat{m}_{q}^{(l_q)}&\triangleq\sum_{p=1}^{l_q}{m_{{q,p}}};\widehat{n}_{q}^{(l_q)}\triangleq\sum_{p=1}^{l_q}{n_{{q,p}}};~~~~~\widehat{u}_{q}^{(l_q)}\triangleq\sum_{p=1}^{l_q}{u_{{q,p}}}.
 \end{align}
By substituting \eqref{eq_Psi_OS_expanded} into \eqref{eq_CDF_OS_Psi} we obtain $F_{\Gamma_{S}}(x)$ in the form of sum-of-products which is suitable to derive the ESR using \eqref{eq_ESR_CDF_eqn} in the next section.
 
% \begin{remark}
% From the left hand side of  \eqref{eq_cross_mul_3_sums}, we notice that second summation depends on the first and the third summation on the first and the second. Application of multinomial theorem for \eqref{eq_cross_mul_3_sums} would have made (\ref{eq_CDF_OS_Psi_beginning}) difficult to track and evaluate in this scenario. Subsequently, evaluating $\Psi(x)^k$ in \eqref{eq_CDF_OS_Psi} using multinomial expansion would have further added  complexity of the closed-form derivation.

Application of the results in \eqref{eq_cross_mul_3_sums} and (\ref{eq_cross_mul_4_sums}) to convert a product of summations to a summation of products via repeated multiplications (as opposed to the method of multinomial expansion) provides a tractable approach which greatly simplifies closed-form ESR analysis. Here we note that both \eqref{eq_cross_mul_3_sums} and (\ref{eq_cross_mul_4_sums}) are required due to the existence of multiple destinations, which increases the problem's difficulty level as compared to the single destination case. 
% Proposed method of \eqref{eq_cross_mul_3_sums} and (\ref{eq_cross_mul_4_sums}) over multinomial theorem used in  \cite{Kim_2015_CPSC,Mallik_TWC_14,other_2017_CPSC} is a significant contribution of our closed-form ESR evaluation technique. 
% However, the application of cross multiplication for expansion of the left side of \eqref{eq_cross_mul_3_sums} and \eqref{eq_cross_mul_4_sums} is straightforward and makes the expression look easy to track and evaluate.
% \end{remark}

\subsection{Evaluation of ESR} 
The ESR is evaluated by substituting \eqref{eq_CDF_OS_Psi} in \eqref{eq_ESR_CDF_eqn} as
\begin{align}\label{eq_ESR_OS_Psi_by_x_1}
     C_{erg} &=\frac{1}{\ln(2)} \sum_{k=1}^{K} \sum_{\lb(\mathbf{l},\mathbf{m},\mathbf{n},\mathbf{u}\rb)\in\mathcal{X}}(-1)^{k+1}\binom{K}{k}\Upsilon\nn\\
     &\times \int_1^\infty\frac{x^{\widetilde{m}^{(k)}-\widetilde{u}^{(k)}-1}\exp({-\frac{\widetilde{l}^{(k)}}{\lambda_{D}}x})}{\prod_{q=1}^{k}\big(x+\frac{\lambda_{D}}{l_q\lambda_{E}}\big)^{M_E+\widehat{n}_{q}^{(l_q)}}}dx\\
     \label{eq_ESR_OS_Psi_by_x}
     &=\frac{1}{\ln(2)} \sum_{k=1}^{K} \sum_{\lb(\mathbf{l},\mathbf{m},\mathbf{n},\mathbf{u}\rb)\in\mathcal{X}}(-1)^{k+1}\binom{K}{k}\Upsilon\nn\\
     & \times \left[J_0^{(k)}+J_1^{(k)}\right],
\end{align}
 where $J_0^{(k)}$ and $J_1^{(k)}$ correspond to the integral with $\widetilde{m}^{(k)}-\widetilde{u}^{(k)} = 0$  and $\widetilde{m}^{(k)}-\widetilde{u}^{(k)} \neq 0$, respectively, and are defined as
\begin{align}
\label{eq_J_0_OS_start_eqn}
J_0^{(k)}&=\int_1^\infty\frac{\exp({-\frac{\widetilde{l}^{(k)}}{\lambda_{D}}x})}{x\Big(\prod\limits_{q=1}^{k}\big(x+\frac{\lambda_{D}}{l_q\lambda_{E}}\big)^{M_E+\widehat{n}_{q}^{(l_q)}}\Big)}dx \textcolor{blue}{,}\\ 
\label{eq_J_1_OS_start_eqn}
J_1^{(k)}&=\int_1^\infty\frac{x^{\widetilde{m}^{(k)}-\widetilde{u}^{(k)}-1}\exp({-\frac{\widetilde{l}^{(k)}}{\lambda_{D}}x})}{\prod_{q=1}^{k}\big(x+\frac{\lambda_{D}}{l_q\lambda_{E}}\big)^{M_E+\widehat{n}_{q}^{(l_q)}}}dx.
\end{align}

\begin{table*}
\begin{align}\label{eq_J_0_OS_sol}
J_0^{(k)} &=A \Gamma\Big(0,\frac{\widetilde{l}^{(k)}}{\lambda_{D}}\Big)+\sum_{i=1}^{\mathcal{I}}\sum_{t=1|q\in\mathcal{Q}_i}^{|\mathcal{Q}_i|M_E+\sum_{q\in\mathcal{Q}_i}\widehat{n}_{q}^{(l_q)}}B_{i,t}\Big(\frac{\widetilde{l}^{(k)}}{\lambda_{D}}\Big)^{t-1}\exp\Big(\frac{\widetilde{l}^{(k)}}{l_q \lambda_{E}}\Big)\Gamma\Big(-t+1,\frac{\widetilde{l}^{(k)}}{\lambda_{D}}\Big(1+\frac{\lambda_D}{l_q \lambda_E}\Big)\Big)\nn\\
 &\quad +\sum_{q\in\bar{\mathcal{Q}}}\sum_{t=1}^{M_E+\widehat{n}_{q}^{(l_q)}}C_{q,t}\Big(\frac{\widetilde{l}^{(k)}}{\lambda_{D}}\Big)^{t-1}\exp\Big(\frac{\widetilde{l}^{(k)}}{l_q \lambda_{E}}\Big)\Gamma\Big(-t+1,\frac{\widetilde{l}^{(k)}}{\lambda_{D}}\Big(1+\frac{\lambda_D}{l_q \lambda_E}\Big)\Big). \\
% \end{align}
% \begin{align}
\label{eq_J_1_1_OS_sol}
   J_1^{(k)} &=\sum_{j=0}^{\widetilde{m}^{k}-\widetilde{u}^{k}-1}\binom{\widetilde{m}^{k}-\widetilde{u}^{k}-1}{j}\nn\Big(-\frac{\lambda_{D}}{l_q\lambda_{E}}\Big)^{\widetilde{m}^{k}-\widetilde{u}^{k}-1-j}\exp\Big({\frac{\widetilde{l}^{(k)}}{l_q\lambda_{E}}}\Big)\nn\\
  &\times
   \Big({\frac{\widetilde{l}^{(k)}}{\lambda_{D}}}\Big)^{-(j-kM_E-\sum_{q=1}^{k}\widehat{n}_{q}^{(l_q)})-1}\Gamma\Big(j-kM_E-\sum_{q=1}^{k}\widehat{n}_{q}^{(l_q)}+1,\frac{\widetilde{l}^{(k)}}{\lambda_D}\Big(1+\frac{\lambda_{D}}{l_q\lambda_{E}}\Big)\Big). \\
% \end{align}
% \begin{align}
\label{eq_J_1_2_OS_sol}
J_1^{(k)} &=\sum_{i=1}^{\mathcal{I}}\sum_{t=1|q\in\mathcal{Q}_i}^{|\mathcal{Q}_i|M_E+\sum_{q\in\mathcal{Q}_i}\widehat{n}_{q}^{(l_q)}}\sum_{j=0}^{\widetilde{m}^{k}-\widetilde{u}^{k}-1} \exp\Big({\frac{\widetilde{l}^{(k)}}{l_q\lambda_{E}}}\Big)\binom{\widetilde{m}^{k}-\widetilde{u}^{k}-1}{j}\Big(-\frac{\lambda_{D}}{l_q\lambda_{E}}\Big)^{\widetilde{m}^{k}-\widetilde{u}^{k}-1-j} B_{i,t}\nn\\
  &\times\Big({\frac{\widetilde{l}^{(k)}}{\lambda_{D}}}\Big)^{-(j-t)-1}\Gamma\Big(j-t+1,\frac{\widetilde{l}^{(k)}}{\lambda_D}\Big(1+\frac{\lambda_{D}}{l_q\lambda_{E}}\Big)\Big) \nn\\
  &+   \sum_{q\in\bar{\mathcal{Q}}}\sum_{t=1}^{M_E+\widehat{n}_{q}^{(l_q)}}\sum_{j=0}^{\widetilde{m}^{k}-\widetilde{u}^{k}-1}\exp\Big({\frac{\widetilde{l}^{(k)}}{l_q\lambda_{E}}}\Big)\binom{\widetilde{m}^{k}-\widetilde{u}^{k}-1}{j}\Big(-\frac{\lambda_{D}}{l_q\lambda_{E}}\Big)^{\widetilde{m}^{k}-\widetilde{u}^{k}-1-j} C_{q,t}\nn\\
 &\times\Big({\frac{\widetilde{l}^{(k)}}{\lambda_{D}}}\Big)^{-(j-t)-1}\Gamma\Big(j-t+1,\frac{\widetilde{l}^{(k)}}{\lambda_D}\Big(1+\frac{\lambda_{D}}{l_q\lambda_{E}}\Big)\Big). 
\end{align}
   \hrule
\end{table*}
Solution of the integrals (\ref{eq_J_0_OS_start_eqn}) and (\ref{eq_J_1_OS_start_eqn}) are carried out in Appendix \ref{appendix1} and \ref{appendix2}, respectively. Finally the results are presented for $J_0^{(k)}$ in  \eqref{eq_J_0_OS_sol}, $J_1^{(k)}$  in \eqref{eq_J_1_1_OS_sol} for $l_q$ being same for all $q\in\{1,\ldots, k\}$ in \eqref{eq_J_1_OS_start_eqn} , and $J_1^{(k)}$ in \eqref{eq_J_1_2_OS_sol} in case any of the $l_q$ are distinct for $q\in\{1,\ldots, k\}$ in \eqref{eq_J_1_OS_start_eqn}. In (\ref{eq_J_0_OS_sol})-(\ref{eq_J_1_2_OS_sol}), $A$, $B_{i,t}$ and $C_{q,t}$ are the partial fraction coefficients specific to $J_0^{(k)}$ and $J_1^{(k)}$. These coefficients can be easily found for a given $K$, $L$, $M_D$, and $M_E$.

From the closed-form ESR presented in \eqref{eq_ESR_OS_Psi_by_x}, it is difficult to interpret how the system parameters $K$, $L$, $M_D$, $M_E$, $\lambda_D$, and $\lambda_E$ affect the ESR; this is due to the presence of the incomplete Gamma function as well as multiple interdependent summations. By looking at $F_{\Gamma_{S}}(x)$ in (\ref{eq_CDF_OS_Psi}) along with (\ref{eq_cross_mul_4_sums}) and (\ref{eq_Psi_OS_expanded}), we observe that the ESR in (\ref{eq_ESR_OS_Psi_by_x}) is also computationally intensive. As $K$, $L$, and $M_D$ (and to a lesser extent, $M_E$) increase, the number of summation and multiplication terms increase significantly. 
A more straightforward ESR expression would provide the required system design parameters quickly when the system is optimized with this metric. Simplified system design is preferable in online systems where latency is a limiting factor, such as delay-sensitive communications in IoT applications. Hence, in the following two sections, we analyze the ESR in the high-SNR regime, and we also find the asymptotic ESR using a less complex expression which can provide further insight. Next, we provide the result for the special case of a single destination. 
\subsubsection{Single Destination Case}\label{sec_OS_Esact_ESR_Single_Dest_case}

For a special case of $L=1$, \eqref{eq_ESR_OS_Psi_by_x} is obtained as
\begin{align}\label{eq_ESR_L_1_OS_Psi_by_x}
  C_{erg} &=\frac{1}{\ln(2)} \sum_{k=1}^{K} (-1)^{k+1}\binom{K}{k}\big[I_0^{(k)}+I_1^{(k)}\big],
\end{align}
where $I_0^{(k)}$ corresponds to the case where $\widehat{n}^{(k)} + u = 0$, and is expressed as
\begin{align}
\label{eq_I_0_L_1_OS}
I_0^{(k)}&=\sum_{\mathbf{m} \in \mathcal{M}^{(k)}} (-1)^{\widehat{m}^{(k)}}\frac{{\lambda_D}^{kM_E-\widehat{m}^{(k)}}\exp\big(\frac{k}{\lambda_{D}}\big)}{{\lambda_E}^{kM_E}\big(\prod\limits_{q=1}^{k}{m_q!}\big)}\nn\\
% &\hspace{1cm}
&\times \bigg[\Gamma\Big(0,\frac{k }{\lambda_{D}}\Big)  -\exp\Big({\frac{k}{\lambda_{E}}}\Big) \sum_{t=1}^{k M_E}\Big(\frac{{k}}{\lambda_{E}}\Big)^{k M_E-t}\nn\\
&\times \Gamma\Big(-\big(k M_E-t\big), \frac{k}{\lambda_{D}}\Big( 1+\frac{\lambda_{D}}{\lambda_{E}}\Big)\Big)\bigg].
\end{align}
Similarly, $I_1^{(k)}$ corresponds to the condition $\widehat{n}^{(k)} + u \neq 0$, and is expressed as
  \begin{align}
  \label{eq_I_1_L_1_OS}  
&I_1^{(k)}=\sum_{\mathbf{m} \in \mathcal{M}^{(k)}}\sum_{\mathbf{n} \in \mathcal{N}(\mathbf{m})}\sum_{u=0}^{\widehat{m}^{(k)}-\widehat{n}^{(k)}}\sum_{j=0}^{\widehat{n}^{(k)}+u-1}(-1)^{\widehat{m}^{(k)}+\widehat{n}^{(k)}+u+j}\nn \\
&\times\frac{\binom{\widehat{m}^{(k)}-\widehat{n}^{(k)}}{u}\binom{\widehat{n}^{(k)}+u-1}{j} \big(\prod\limits_{q=1}^{k}\binom{m_q}{n_q}{\Gamma\left(M_E+n_q \right)}\big)}{k^{u-kM_E-j}{\lambda_{D}}^{\widehat{m}^{(k)}-\widehat{n}^{(k)}-u}{\lambda_E}^{kM_E+j}\big(\prod\limits_{q=1}^{k}{m_q!}\big){\Gamma(M_E)}^k}\nn\\
&\times\Gamma\big(-\left(k M_E-u+j\right),\frac{k}{\lambda_{D}}\big( 1+\frac{\lambda_{D}}{\lambda_{E}}\big)\big)\exp\big({\frac{k}{\lambda_{D}}}+{\frac{k}{\lambda_{E}}}\big).
  \end{align}
% \begin{align}
% \label{eq_I_1_L_1_OS}
% &I_1^{(k)}=\sum_{\mathbf{m} \in \mathcal{M}^{(k)}}\sum_{\mathbf{n} \in \mathcal{N}(\mathbf{m})}\sum_{u=0}^{\widehat{m}^{(k)}-\widehat{n}^{(k)}}\sum_{j=0}^{\widehat{n}^{(k)}+u-1}(-1)^{\widehat{m}^{(k)}+\widehat{n}^{(k)}+u+j}\nn \\
% &\times\frac{\binom{\widehat{m}^{(k)}-\widehat{n}^{(k)}}{u}\binom{\widehat{n}^{(k)}+u-1}{j} \big(\prod\limits_{q=1}^{k}\binom{m_q}{n_q}{\Gamma\left(M_E+n_q \right)}\big)}{k^{u-kM_E-j}{\lambda_{D}}^{\widehat{m}^{(k)}-\widehat{n}^{(k)}-u}{\lambda_E}^{kM_E+j}\big(\prod\limits_{q=1}^{k}{m_q!}\big){\Gamma(M_E)}^k}\nn\\
% &\times\Gamma\big(-\left(k M_E-u+j\right),\frac{k}{\lambda_{D}}\big( 1+\frac{\lambda_{D}}{\lambda_{E}}\big)\big)\exp\big({\frac{k}{\lambda_{D}}}+{\frac{k}{\lambda_{E}}}\big).
% \end{align}
Thus, when $L=1$, partial fraction coefficients are available in closed form, and by employing those coefficients the exact ESR  is expressed in (\ref{eq_ESR_L_1_OS_Psi_by_x}) which is much simpler than the corresponding expression (\ref{eq_ESR_OS_Psi_by_x}) for multiple destination case.

\subsection{High-SNR Approximation}\label{section_High_SNR_Approximation}

To reduce the computational complexity of the solution and to assess more easily the behavior of the ESR with the system parameters $K$, $L$, $M_D$, $M_E$, $\lambda_D$, and $\lambda_E$, in this subsection, we determine an approximate ESR expression at high SNR. This high-SNR approximation also helps us find the asymptotic ESR in the next section.

By the ``high-SNR'' condition, we mean that $\gamma_D^{(k,l)}>>1$ and $\gamma_E^{(k)}>>1$ for all $k$  and $l$. Therefore, we use the approximation $\Gamma_S^{(k,l)}=\frac{1+\max\limits_{\forall l}\left\{\gamma_{D}^{(k,l)}\right\}}{1+\gamma_{E}^{(k)}} \approx \frac{\max\limits_{\forall l}\left\{\gamma_{D}^{(k,l)}\right\}}{\gamma_{E}^{(k)}}$ in \eqref{eq_Gamma_s}. With this approximation, the high-SNR CDF of $\Gamma_S$ in (\ref{eq_CDF_OS_Psi_beginning_other}) is then obtained as 
\begin{align}\label{eq_CDF_OS_High_SNR}
 F_{\Gamma_{S}}(x)& =\mathbb P\lb[\Gamma_{S} \le x\rb]\nn\\
 &= \lb[\int_0^\infty F_{\max\limits_{\forall l}\{ \gamma_{D}^{(k,l)}\}}\left(xy\right)f_{\gamma_{E}^{(k)}}(y) dy\rb]^{K}.
\end{align}
The integral solution above is expressed exactly as in (\ref{eq_CDF_OS_Psi}), where
\begin{align}\label{eq_Psi_OS_High_SNR}
\Psi(x)&=\sum_{l=1}^{L}\sum_{\mathbf{m} \in \mathcal{M}^{(l)}}(-1)^{l+1}\binom{L}{l}\nn\\
&\times \frac{{\lambda_D}^{M_E}\Gamma\left(M_E+\widehat{m}^{(l)}\right)x^{\widehat{m}^{(l)}}}{{\lambda_E}^{M_E}l^{M_E+\widehat{m}^{(l)}}\Big(\prod\limits_{p=1}^{l}m_{p}!\Big)\Gamma(M_E)\big(x+\frac{\lambda_{D}}{l\lambda_{E}}\big)^{M_E+\widehat{m}^{(l)}}}.
\end{align}
This expression is obtained using manipulations similar to those used in the derivation of \eqref{eq_Psi_OS_before_expansion} in the exact ESR analysis. 
As we did in \eqref{eq_Psi_OS_expanded}, we convert $\Psi^k(x)$ from a product-of-sums into a sum-of-products form to evaluate the ESR, yielding
%\vspace{-0.2cm}
\begin{align}\label{eq_Psi_OS_High_SNR_expanded}
   \Psi^k(x)&= \sum_{\mathbf{l} \in \mathcal{L}^{(k)}}\sum_{\mathbf{m}\in\mathcal{M}^{(k,l_k)}}\Upsilon\frac{x^{\widetilde{m}^{(k)}}}{\prod\limits_{q=1}^{k}\big(x+\frac{\lambda_{D}}{l_q\lambda_{E}}\big)^{M_E+\widehat{m}_{{q}}^{(l_q)}}},
\end{align}
where 
\begin{align}  
% \widetilde{\sum}&\triangleq\sum_{\mathbf{l} \in \mathcal{L}^{(k)}}\sum_{\mathbf{m}\in\mathcal{M}^{(k,l_k)}},\nn\\
\Upsilon&\triangleq\prod\limits_{q=1}^{k}(-1)^{l_q+1}\binom{L}{l_q} \frac{{\lambda_D}^{M_E}\Gamma\left(M_E+\widehat{m}_{q}^{(l_q)}\right)}{{\lambda_E}^{M_E}{l_q}^{M_E+\widehat{m}_{q}^{(l_q)}}\Big(\prod\limits_{p=1}^{l_q}m_{{q,p}}!\Big){\Gamma(M_E)}},
\end{align}

with the definitions already provided in \eqref{eq_definition_OS_2}.
% and
% \begin{align}\label{eq_definition_OS_HSNR_2}
% \widehat{m}_{q}^{(l_q)}\triangleq\sum_{p=1}^{l_q}{m_{{q,p}}}.
% \end{align}
By substituting \eqref{eq_Psi_OS_High_SNR_expanded} in \eqref{eq_CDF_OS_Psi} and then using \eqref{eq_ESR_CDF_eqn}, the ESR in \eqref{eq_ESR_OS_Psi_by_x_1} becomes, at high SNR, 
\begin{align}\label{eq_ESR_OS_High_SNR_Psi_by_x}
     C_{erg} &=\frac{1}{\ln(2)} \sum_{k=1}^{K}\sum_{\mathbf{l} \in \mathcal{L}^{(k)}}\sum_{\mathbf{m}\in\mathcal{M}^{(k,\mathbf{l})}}(-1)^{k+1}\binom{K}{k}\Upsilon\nn\\
     & \times \left[J_0^{(k)}+J_1^{(k)}\right].
\end{align}
Similar to \eqref{eq_ESR_OS_Psi_by_x}, 
% the integral above can be divided into $J_0^{(k)}$ and $J_1^{(k)}$, where
$J_0^{(k)}$ corresponds to the case when $\widetilde{m}^{(k)}= 0$ and $J_1^{(k)}$ corresponds to the case when $\widetilde{m}^{(k)}\ne 0$. $J_0^{(k)}$ and $J_1^{(k)}$ are expressed as
\begin{align}
\label{high_snr_J0}
J_0^{(k)}&=\int_1^{\infty}\frac{1}{x\Big(\prod\limits_{q=1}^{k}\big(x+\frac{\lambda_{D}}{l_q\lambda_{E}}\big)^{M_E}\Big)}dx ,\\
\label{high_snr_J1}
J_1^{(k)}&=\int_1^{\infty}\frac{x^{\widetilde{m}^{(k)}-1}}{\prod\limits_{q=1}^{k}\big(x+\frac{\lambda_{D}}{l_q\lambda_{E}}\big)^{M_E+\widehat{m}_{{q}}^{(l_q)}}}dx.
\end{align}
The solution approach of $J_0^{(k)}$ at high SNR in (\ref{high_snr_J0}) is similar to that of $J_0^{(k)}$ in (\ref{eq_J_0_OS_solution}) following the partial fraction method. The solution is presented in \eqref{eq_J_0_High_SNR_OS_sol_final}. 
% with $A_0=A_1=\sum_{r}A_{r,1}=\prod_{q=1}^{k}\big(l_q\lambda_E/\lambda_D\big)^{M_E}$, and  $A_t=\prod_{q=1}^{k}\big(l_q\lambda_E/\lambda_D\big)^{M_E-t+1}$.
% The solution of $J_0^{(k)}$ is given in \eqref{eq_I_0_OS_High_SNR}.
The solution of the integral (\ref{high_snr_J1}) is presented in Appendix \ref{appendix4}. An expression for $J_1^{(k)}$ is presented in \eqref{eq_J_1_1_High_SNR_OS_sol_final} for the case where $l_q$ are equal for all $q\in\{1,\ldots, k\}$, and in \eqref{eq_J_1_2_High_SNR_OS_sol_final} for the case where any of the $l_q$ are unequal for $q\in\{1,\ldots, k\}$ in \eqref{high_snr_J1}. 
$B_{i,t}$ and $C_{q,t}$ are the partial fraction coefficients specific to $J_0^{(k)}$ and $J_1^{(k)}$ at high SNR and can be easily obtained for a given $K$, $L$, $M_D$, and $M_E$.

\begin{table*}
\begin{align} \label{eq_J_0_High_SNR_OS_sol_final}
    J_0^{(k)}&=\sum_{i=1}^{\mathcal{I}}\sum_{t=1|q\in\mathcal{Q}_i}^{1}{B_{i,t}}\ln\Big(1+\frac{\lambda_{D}}{l_q\lambda_{E}}\Big)+\sum_{i=1}^{\mathcal{I}}\sum_{t=2|q\in\mathcal{Q}_i}^{|\mathcal{Q}_i| M_E}\frac{B_{i,t}}{(t-1)\big(1+\frac{\lambda_{D}}{l_q\lambda_{E}}\big)^{t-1}}\nn\\
 &+\sum_{q\in\bar{\mathcal{Q}}}\sum_{t=1}^{1}{C_{q,t}}\ln\Big(1+\frac{\lambda_{D}}{l_q\lambda_{E}}\Big)+\sum_{q\in\bar{\mathcal{Q}}}\sum_{t=2}^{M_E}\frac{C_{q,t}}{(t-1)\big(1+\frac{\lambda_{D}}{l_q\lambda_{E}}\big)^{t-1}}.\\
 \label{eq_J_1_1_High_SNR_OS_sol_final}
J_1^{(k)}&=\sum_{j=0}^{\widetilde{m}^{(k)}-1}\binom{\widetilde{m}^{(k)}-1}{j}\Big(-\frac{\lambda_{D}}{l_q\lambda_{E}}\Big)^{\widetilde{m}^{(k)}-1-j}\frac{\big(1+\frac{\lambda_D}{l_q\lambda_E}\big)^{-(kM_E+\widetilde{m}^{(k)}-1-j)}}{kM_E+\widetilde{m}^{(k)}-1-j}.\\
\label{eq_J_1_2_High_SNR_OS_sol_final}
J_1^{(k)}&=\sum_{i=1}^{\mathcal{I}}\sum_{t=1}^{1}-{B_{i,t}}\ln{\Big(1+\frac{\lambda_{D}}{l_q\lambda_{E}}\Big)}+\sum_{i=1}^{\mathcal{I}}\sum_{t=2|q\in\mathcal{Q}_i}^{|\mathcal{Q}_i| M_E+\sum_{q\in\mathcal{Q}_i}\widehat{m}_{q}^{l_q}}\frac{B_{i,t}}{(t-1)\big(1+\frac{\lambda_{D}}{l_q\lambda_{E}}\big)^{t-1}}\nn\\
 &+\sum_{q\in\bar{\mathcal{Q}}}\sum_{t=1}^{1}-{C_{q,t}}\ln{\Big(1+\frac{\lambda_{D}}{l_q\lambda_{E}}\Big)}+\sum_{q\in\bar{\mathcal{Q}}}\sum_{t=2}^{M_E+\widehat{m}_{q}^{l_q}}\frac{C_{q,t}}{(t-1)\big(1+\frac{\lambda_{D}}{l_q\lambda_{E}}\big)^{t-1}}.
\end{align}
\hrule
\end{table*}

By comparing the exact ESR in \eqref{eq_ESR_OS_Psi_by_x} and the high-SNR approximate ESR in  \eqref{eq_ESR_OS_High_SNR_Psi_by_x}, we observe that evaluating the exact ESR is more computationally intensive than the high-SNR ESR as the summations over $\mathcal{N}^{(k,\mathbf{l})}(\mathbf{m})$ and $\mathcal{U}^{(k,\mathbf{l})}(\mathbf{m},\mathbf{n})$ are absent in \eqref{eq_ESR_OS_Psi_by_x_1}. 
To obtain insights and investigate computational complexity advantage in the high-SNR regime,  we focus on the special case of a single destination in the following analysis.

% \textcolor{blue}{The computation of the exact ESR in \eqref{eq_ESR_OS_Psi_by_x} using $\Psi^k(x)$ in \eqref{eq_Psi_OS_before_expansion} with the help of \eqref{eq_cross_mul_3_sums} requires $\sum_{k=1}^{K}\sum_{l=1}^{L}(M_D(M_D+1)(M_D+2)/6)^{kl}$ additions, which is in the order of $\sum_{k=1}^{K}\sum_{l=1}^{L}({M}_{D}^3/6)^{kl}\approx ({M}_{D}^3/6)^{KL}$. For the  $k^{\text{th}}$ and $l^{\text{th}}$ summation iteration, there are $kl$ multiplications, therefore the number of multiplications required is in the order of  $\sum_{k=1}^{K}\sum_{l=1}^{L}kl({M}_{D}^3/6)^{kl}\approx KL({M}_{D}^3/6)^{KL}$. 
% The computation of the high-SNR ESR in \eqref{eq_ESR_OS_High_SNR_Psi_by_x} using $\Psi^k(x)$ in \eqref{eq_Psi_OS_High_SNR_expanded}  requires $\sum_{k=1}^{K}\sum_{l=1}^{L}M_D^{kl}$ additions which is in the order of $M_D^{KL}$.   Also, as $k$ multiplications are required in the $k^{\text{th}}$ and $l^{\text{th}}$ summation iteration, the number of multiplications is in the order of $KL(M_D)^{KL}$. Clearly, the high-SNR approximation has less computation.}

\subsubsection{High-SNR Approximation in Single Destination Case}\label{sec_OS_HSNR_ESR_Single_Dest_case}

For the special case of $L=1$, \eqref{eq_ESR_OS_High_SNR_Psi_by_x} has the form of \eqref{eq_ESR_L_1_OS_Psi_by_x}, where  

\begin{align}\label{eq_I_0_L_1_OS_High_SNR}
%   J_0^{(k)}&=
%     \Big(\frac{\lambda_{D}}{\lambda_{E}}\Big)^{-kM_E}\ln\Big(1+\frac{\lambda_{D}}{\lambda_{E}}\Big)-\sum_{t=2}^{kM_E}\frac{\big(\frac{\lambda_{D}}{\lambda_{E}}\big)^{-kM_E+t-1}}{(t-1)\big(1+\frac{\lambda_{D}}{\lambda_{E}}\big)^{t-1}},\nn\\
     I_0^{(k)}&=\ln \Big(1+ \frac{\lambda_{D}}{\lambda_{E}}\Big)\nn\\
     &-\sum_{t=1}^{k M_E-1}\frac{\big(\frac{\lambda_D}{\lambda_E}\big)^{kM_E-t}}{\big(1+\frac{\lambda_{D}}{\lambda_{E}} \big)^{kM_E-t}\left(kM_E-t\right)},
\end{align}

and
\begin{align}\label{eq_I_1_L_1_OS_High_SNR}
% I_1^{(k)}=&\sum_{j=0}^{\widehat{m}^{(k)}-1}{(-1)^{\widehat{m}^{(k)}-j-1}}\binom{\widehat{m}^{(k)}-1}{j}\frac{\big(\frac{\lambda_D}{\lambda_E}\big)^{(\widehat{m}^{(k)}-j-1)}\big(1+\frac{\lambda_{D}}{\lambda_{E}} \big)^{-\left(kM_E+\widehat{m}^{(k)}-j-1\right)}}{\left(kM_E+\widehat{m}^{(k)}-j-1\right)},
I_1^{(k)}&=\sum_{\mathbf{m} \in \mathcal{M}^{(k)}}\sum_{j=0}^{\widehat{m}^{(k)}-1}{(-1)^{\widehat{m}^{(k)}-j-1}}\binom{\widehat{m}^{(k)}-1}{j}\nn\\
&\times \frac{\big(\frac{\lambda_D}{\lambda_E}\big)^{(kM_E+\widehat{m}^{(k)}-j-1)}\big(\prod\limits_{q=1}^{k}{\Gamma\left(M_E+m_q\right)} \big)}{\left(1+\frac{\lambda_{D}}{\lambda_{E}} \right)^{\left(kM_E+\widehat{m}^{(k)}-j-1\right)}\big(\prod\limits_{q=1}^{k}{m_q!}\big){\Gamma(M_E)}^k}\nn\\
&\times \frac{1}{\left(kM_E+\widehat{m}^{(k)}-j-1\right)}.
\end{align}
We notice that the ESR in (\ref{eq_ESR_OS_Psi_by_x}) with high-SNR approximation via \eqref{eq_I_0_L_1_OS_High_SNR} and \eqref{eq_I_1_L_1_OS_High_SNR} is a summation of two components. One component varies as a logarithmic function of the ratio $\lambda_D/\lambda_E$ and is independent of $K$, $M_D$, and $M_E$. The other component tends to be a constant as $\lambda_D/\lambda_E$ grows asymptotically; however, it depends on $K$, $M_D$, and $M_E$. This will be made clearer when we present the asymptotic analysis in the next section.

If we compare the computational complexity in terms of the number of addition and multiplication operations required to compute the exact ESR for the single destination case, it can be seen that the high-SNR approximate expression is less computationally intensive. This is due to the fact that the CDF of $\Gamma_{S}^{(k)}$ in \eqref{eq_CDF_OS_Psi} has a smaller number of summation terms in the high-SNR approximation scenario.  For $L=1$, the computation of the CDF in \eqref{eq_CDF_OS_Psi} for the exact ESR using $\Psi^k(x)$ in \eqref{eq_Psi_OS_before_expansion} with the help of \eqref{eq_cross_mul_3_sums} requires $\sum_{k=1}^K(M_D(M_D+1)/2)^k$ additions, which is in the order of $\sum_{k=1}^{K}(M_D/\sqrt{2})^{2k}\approx (M_D/\sqrt{2})^{2K}$. For the  $k^{\text{th}}$ summation iteration, there are $k$ multiplications, therefore the number of multiplications required is in the order of  $\sum_{k=1}^{K}(k(M_D/\sqrt{2})^{2k})\approx (K(M_D/\sqrt{2})^{2K})$. 
The computation of the CDF in \eqref{eq_CDF_OS_Psi} for the high-SNR approximation using $\Psi^k(x)$ in \eqref{eq_Psi_OS_High_SNR_expanded}  requires $\sum_{k=1}^KM_D^k$ additions. Hence, it is in the order of $M_D^{K}$.   Also, as $k$ multiplications are required in the $k^{\text{th}}$ summation iteration, the number of multiplications is in the order of $KM_D^K$. Clearly, the high-SNR approximation has lower computational complexity.

\subsection{Asymptotic Analysis}\label{subsection_Asymptotic_Analysis_of_ESR_in_OS}
% %\vspace{-0.3cm}
To provide better insight regarding how the ESR depends on the system parameters $K$, $L$, $M_D$, $M_E$, $\lambda_D$, and $\lambda_E$, we perform asymptotic analysis by assuming $\lambda_D \rightarrow{\infty}$ for a given $\lambda_E$ in the high-SNR approximate ESR expression derived in (\ref{eq_ESR_OS_High_SNR_Psi_by_x}). This assumption is based on a practical scenario where the destinations are closer to the transmitters than the eavesdropper. With this assumption, using the approximation $1+\frac{\lambda_{D}}{l_q\lambda_{E}} \approx \frac{\lambda_{D}}{l_q\lambda_{E}}$ for each $q \in \{1 ,\ldots, k\}$ in $J_0^{(k)}$ given by \eqref{eq_J_0_High_SNR_OS_sol_final} we find 
\begin{align}\label{eq_J_0_asympt_OS}
    % J_0^{(k)}
    % &=\lim\limits_{1+\frac{\lambda_D}{\lambda_E} \rightarrow \frac{\lambda_D}{\lambda_E}}\int_1^{\infty}\frac{1}{x\Big(\prod\limits_{q=1}^{k}\big(x+\frac{\lambda_{D}}{l_q\lambda_{E}}\big)^{M_E}\Big)}dx \nonumber\\
%   &=A_1 \ln\big(\frac{\lambda_{D}}{l_i\lambda_{E}}\big)+\sum_{r}A_{r,1} \ln\big(\frac{\lambda_{D}}{l_r\lambda_{E}}\big)+\sum_{t=2}^{\vartheta M_E}\frac{A_{t}}{(t-1)\big(\frac{\lambda_{D}}{l_i\lambda_{E}}\big)^{t-1}}+\sum_{r}\sum_{t_r=2}^{M_E}\frac{A_{r,t_r}}{(t_r-1)\big(\frac{\lambda_{D}}{l_r\lambda_{E}}\big)^{t_r-1}}\nn\\
&J_0^{(k)}=\sum_{i=1}^{\mathcal{I}}\sum_{t=1|q\in\mathcal{Q}_i}^{1}{B_{i,t}}\ln\Big(\frac{\lambda_{D}}{l_q\lambda_{E}}\Big)\nn\\
&+\sum_{i=1}^{\mathcal{I}}\sum_{t=2|q\in\mathcal{Q}_i}^{|\mathcal{Q}_i| M_E}\frac{B_{i,t}}{(t-1)\big(\frac{\lambda_{D}}{l_q\lambda_{E}}\big)^{t-1}}\nn\\
 &+\sum_{q\in\bar{\mathcal{Q}}}\sum_{t=1}^{1}{C_{q,t}}\ln\Big(\frac{\lambda_{D}}{l_q\lambda_{E}}\Big)+\sum_{q\in\bar{\mathcal{Q}}}\sum_{t=2}^{M_E}\frac{C_{q,t}}{(t-1)\big(\frac{\lambda_{D}}{l_q\lambda_{E}}\big)^{t-1}}.
\end{align}
Using the same approximation in \eqref{eq_int_J_1_High_SNR_OS_appendix_1}, we obtain $J_1^{(k)}$ 
when all $l_q$ are equal for all $q\in\{1 ,\ldots, k\}$ as

\begin{align}\label{eq_J_1_asympt_OS_1}
%   J_1^{(k)}&=\lim\limits_{1+\frac{\lambda_D}{\lambda_E} \rightarrow \frac{\lambda_D}{\lambda_E}}\int_1^{\infty}\frac{x^{\widetilde{m}_{{q,p}}^{(l_q)}-1}}{\prod\limits_{q=1}^{k}\big(x+\frac{\lambda_{D}}{l_q\lambda_{E}}\big)^{M_E+\widetilde{m}_{{q,p}}^{(l_q)}}}dx\nn\\
% &=
% \begin{cases}
%   \sum_{v=0}^{\widetilde{m}_{{q,p}}^{(l_q)}-1}\binom{\widetilde{m}_{{q,p}}^{(l_q)}-1}{v}\big(-\frac{\lambda_{D}}{l_i\lambda_{E}}\big)^{\widetilde{m}_{{q,p}}^{(l_q)}-1-v}\nn\\
%   \frac{\big(\frac{\lambda_D}{l_i\lambda_E}\big)^{v-kM_E-\widetilde{m}_{{q,p}}^{(l_i)}+1}}{kM_E+\widetilde{m}_{{q,p}}^{(l_i)}-1-v} &, l_i = l_j, \forall\{i,j\}\in q \\
%   B_1\ln{\big(\frac{\lambda_{D}}{l_i\lambda_{E}}\big)}+\sum_{r}B_{r,1}\ln{\big(\frac{\lambda_{D}}{l_r\lambda_{E}}\big)}\nn\\
%   +\sum_{t=2}^{\vartheta M_E+\sum_{q=1}^{\vartheta}\sum_{p=1}^{l_q}{m_{{q,p}}}}\frac{B_t}{(t-1)\big(\frac{\lambda_{D}}{l_i\lambda_{E}}\big)^{t-1}}\nn\\
%   + \sum_{r}\sum_{t_r=2}^{M_E+\widehat{m}_{{r,p}}^{(l_r)}}\frac{B_{r,t_r}}{(v-1)\big(x+\frac{\lambda_{D}}{l_r\lambda_{E}}\big)^{v-1}} &, \text{otherwise}
%     \end{cases}.\\
J_1^{(k)}
% &=\int_1^{\infty}\frac{x^{\widetilde{m}^{(k)}-1}}{\big(x+\frac{\lambda_{D}}{l_q\lambda_{E}}\big)^{k M_E+\widetilde{m}^{(k)}}}dx\nn\\
&=\sum_{j=0}^{\widetilde{m}^{(k)}-1}(-1)^{\widetilde{m}^{(k)}-1-j}\binom{\widetilde{m}^{(k)}-1}{j}\nn\\
&\times \frac{\big(\frac{\lambda_D}{l_q\lambda_E}\big)^{-kM_E}}{kM_E+\widetilde{m}^{(k)}-1-j},
\end{align} and 
when any of the $l_q$ are unequal for $q\in\{1 ,\ldots, k\}$ in (\ref{eq_int_J_1_High_SNR_OS_appendix_2}) as
\begin{align}\label{eq_J_1_asympt_OS_2}
%   J_1^{(k)}&=\lim\limits_{1+\frac{\lambda_D}{\lambda_E} \rightarrow \frac{\lambda_D}{\lambda_E}}\int_1^{\infty}\frac{x^{\widetilde{m}_{{q,p}}^{(l_q)}-1}}{\prod\limits_{q=1}^{k}\big(x+\frac{\lambda_{D}}{l_q\lambda_{E}}\big)^{M_E+\widetilde{m}_{{q,p}}^{(l_q)}}}dx\nn\\
% &=
% \begin{cases}
%   \sum_{v=0}^{\widetilde{m}_{{q,p}}^{(l_q)}-1}\binom{\widetilde{m}_{{q,p}}^{(l_q)}-1}{v}\big(-\frac{\lambda_{D}}{l_i\lambda_{E}}\big)^{\widetilde{m}_{{q,p}}^{(l_q)}-1-v}\nn\\
%   \frac{\big(\frac{\lambda_D}{l_i\lambda_E}\big)^{v-kM_E-\widetilde{m}_{{q,p}}^{(l_i)}+1}}{kM_E+\widetilde{m}_{{q,p}}^{(l_i)}-1-v} &, l_i = l_j, \forall\{i,j\}\in q \\
%   B_1\ln{\big(\frac{\lambda_{D}}{l_i\lambda_{E}}\big)}+\sum_{r}B_{r,1}\ln{\big(\frac{\lambda_{D}}{l_r\lambda_{E}}\big)}\nn\\
%   +\sum_{t=2}^{\vartheta M_E+\sum_{q=1}^{\vartheta}\sum_{p=1}^{l_q}{m_{{q,p}}}}\frac{B_t}{(t-1)\big(\frac{\lambda_{D}}{l_i\lambda_{E}}\big)^{t-1}}\nn\\
%   + \sum_{r}\sum_{t_r=2}^{M_E+\widehat{m}_{{r,p}}^{(l_r)}}\frac{B_{r,t_r}}{(v-1)\big(x+\frac{\lambda_{D}}{l_r\lambda_{E}}\big)^{v-1}} &, \text{otherwise}
%     \end{cases}.\\
&J_1^{(k)}
=\sum_{i=1}^{\mathcal{I}}\sum_{t=1}^{1}-{B_{i,t}}\ln{\Big(\frac{\lambda_{D}}{l_q\lambda_{E}}\Big)}\nn\\
&+\sum_{i=1}^{\mathcal{I}}\sum_{t=2|q\in\mathcal{Q}_i}^{|\mathcal{Q}_i| M_E+\sum_{q\in\mathcal{Q}_i}\widehat{m}_{q}^{l_q}}\frac{B_{i,t}}{(t-1)\big(\frac{\lambda_{D}}{l_q\lambda_{E}}\big)^{t-1}}\nn\\
 &+\sum_{q\in\bar{\mathcal{Q}}}\sum_{t=1}^{1}-{C_{q,t}}\ln{\Big(\frac{\lambda_{D}}{l_q\lambda_{E}}\Big)}+\sum_{q\in\bar{\mathcal{Q}}}\sum_{t=2}^{M_E+\widehat{m}_{q}^{l_q}}\frac{C_{q,t}}{(t-1)\big(\frac{\lambda_{D}}{l_q\lambda_{E}}\big)^{t-1}}.
\end{align}
The asymptotic ESR $C_{erg}^{\infty}$ is obtained by substituting \eqref{eq_J_0_asympt_OS}, \eqref{eq_J_1_asympt_OS_1} and \eqref{eq_J_1_asympt_OS_2} into \eqref{eq_ESR_OS_High_SNR_Psi_by_x}. Next, we analyze the special case of a single destination for more insights.
\subsubsection{Asymptotic Analysis in Single Destination Case} By substituting $1+\frac{\lambda_{D}}{\lambda_{E}} \approx \frac{\lambda_{D}}{\lambda_{E}}$ in
% \eqref{eq_I_0_L_1_OS_High_SNR} and \eqref{eq_I_1_L_1_OS_High_SNR}, then using these in
\eqref{eq_I_0_L_1_OS_High_SNR} and after performing some mathematical simplifications, 
we find 
\begin{align}\label{eq_C_0_asympt_OS}
    I_0^{(k)}&  =\ln\Big(\frac{\lambda_D}{\lambda_E}\Big)-\sum_{i=1}^{kM_E-1}\frac{1}{kM_E-i}
    \nn\\
    &
    =\ln\Big(\frac{\lambda_D}{\lambda_E}\Big)-H_{kM_E-1},
\end{align}
where $H_{N}$ is the $N^{\text{th}}$ harmonic number.
Using the same approximation in \eqref{eq_I_1_L_1_OS_High_SNR}, and after writing $\binom{\widehat{m}^{(k)}-1}{j}$ in the factorial form, we obtain $I_1^{(k)}$ as
\begin{align}
\label{eq_C_1_asympt_OS_append}
 & I_1^{(k)}
  =\sum_{\mathbf{m} \in \mathcal{M}^{(k)}}\frac{\big(\prod\limits_{q=1}^{k}{\Gamma\left(M_E+m_q\right)} \big)}{\big(\prod\limits_{q=1}^{k}m_q!\big){\Gamma(M_E)}^k}
  \nn\\
  &  \times 
  \sum_{u=0}^{\widehat{m}^{(k)}-1}\frac{(-1)^{\widehat{m}^{(k)}-u-1}\Gamma(\widehat{m}^{(k)})}{u!(\widehat{m}^{(k)}-u-1)!\left(kM_E+\widehat{m}^{(k)}-u-1\right)}\\
   \label{eq_Xi_first_time}
 &=\sum_{\mathbf{m} \in \mathcal{M}^{(k)}}\frac{\big(\prod\limits_{q=1}^{k}{\Gamma\left(M_E+m_q\right)} \big)\Gamma(\widehat{m}^{(k)})(k M_E-1)!}{\big(\prod\limits_{q=1}^{k}m_q!\big){\Gamma(M_E)}^k(kM_E+\widehat{m}^{(k)}-1)!} \Xi,
    % &\times\sum_{v=0}^{\widehat{m}^{(k)}-1} \frac{(-1)^{\widehat{m}^{(k)}-v-1}\left(\prod_{l=0,l \neq v}^{\widehat{m}^{(k)}-1}l!(\widehat{m}^{(k)}-l-1)!(kM_E+\widehat{m}^{(k)}-l-1)\right)}{\left({\prod_{l=0}^{\widehat{m}^{(k)}-1}l!(\widehat{m}^{(k)}-l-1)!}\right)}\\
% &=\sum_{\mathbf{m} \in \mathcal{M}^{(k)}}\frac{\big(\prod_{i=1}^{k}{\Gamma\left(M_E+m_q\right)} \big)\Gamma(\widehat{m}^{(k)})\Gamma(kM_E)}{\Gamma(kM_E+\widehat{m}^{(k)})\big(\prod_{q=1}^{k}m_q!\big){\Gamma(M_E)}^k}.
\end{align}
where 
\begin{align}
    &\Xi=\sum_{v=0}^{\widehat{m}^{(k)}-1}(-1)^{\widehat{m}^{(k)}-v-1} 
    \nn\\
    &\times
    \frac{\prod \limits_{u=0,u \neq v}^{\widehat{m}^{(k)}-1}\big(u!(\widehat{m}^{(k)}-u-1)!(kM_E+\widehat{m}^{(k)}-u-1)\big)}{\prod\limits_{u=0}^{\widehat{m}^{(k)}-1}\big(u!(\widehat{m}^{(k)}-u-1)!\big)}
    % \nn\\
    % =&
    =1.
\end{align}
The expression (\ref{eq_Xi_first_time}) above is obtained after multiplying both the numerator and denominator of (\ref{eq_C_1_asympt_OS_append}) by $(k M_E-1)!$ and performing some mathematical simplifications. The proof that $\Xi=1$ is straightforward and is omitted due to space constraints.

The asymptotic ESR, $C_{erg}^{\infty}$, % with the high-SNR approximation 
is obtained by substituting \eqref{eq_C_0_asympt_OS} and \eqref{eq_Xi_first_time} into \eqref{eq_ESR_OS_Psi_by_x}. 
% \begin{align}\label{eq_C_erg_Asympt_OS_cases}
%     &C_{erg}^{\infty}=\frac{1}{\ln(2)} \sum_{k=1}^{K}(-1)^{k+1}\binom{K}{k}\big[{C_0^\infty}+{C_1^\infty}\big].
%     % &=\sum_{k=1}^{K}W_k\Big[\underbrace{C_{0,k}^\infty}_{  \left(\widehat{m}^{(k)}-1\right)<0}+\underbrace{C_{1,k}^\infty}_{\left(\widehat{m}^{(k)}-1\right)\geq 0}\Big]\nn\\
%     % &=\sum_{k=1}^{K}W_k\Big[\sum_{m=0}^{M_D-1}w_{0,m,k}\underbrace{C_{0,m,k}^\infty}_{  \left(\widehat{m}^{(k)}-1\right)<0}+\sum_{m=0}^{M_D-1}w_{1,m,k}\underbrace{C_{1,m,k}^\infty}_{\left(\widehat{m}^{(k)}-1\right)\geq 0}\Big]\nn\\
% \end{align}
% where the term $C_0^\infty$ and $C_1^\infty$ correspond to the case $=0$ and $\widehat{m}^{(k)}\geq 1$ respectively. 
After a few simplifications, $C_{erg}^{\infty}$ is expressed as
\begin{align}\label{eq_C_erg_L_1_OS_Asympt}
    %  &C_{erg}^{\infty} =\frac{1}{\ln(2)} \sum_{k=1}^{K}(-1)^{k+1}\binom{K}{k}     %  \nonumber\\
    % %     & \times 
    %     \Big[\ln\Big (\frac{\lambda_{D}}{\lambda_{E}}\Big)-\psi_\infty\Big]\nonumber\\
    &C_{\mathrm{erg}}^{\infty} =\log_2 \Big(\frac{\lambda_{D}}{\lambda_E}\Big)
    % \nonumber\\
    % &
    -\frac{1}{\ln(2)}\sum_{k=1}^{K}(-1)^{k+1}\binom{K}{k} \psi_{\infty}^{(k)},
\end{align}
where 
% \begin{align}\label{eq_psi_OS_Asympt}
%     % \psi_{\infty}= 
%     %  \begin{cases}
%     %         &\sum_{i=1}^{k M_E-1}\frac{1}{\left(k M_E-i\right)}+\sum_{\mathbf{m} \in \mathcal{M}^{(k)}}\sum_{l=0}^{\widehat{m}^{(k)}-1}(-1)^{\widehat{m}^{(k)}-l}\binom{\widehat{m}^{(k)}-1}{l}\left(\prod_{i=1}^{k}\frac{\Gamma\left(M_E+m_q\right)}{\Gamma\left(M_E\right)}\left(\frac{1}{m_q!}\right)\right) \left(\frac{1}{kM_E+\widehat{m}^{(k)}-l-1} \right) ,    \left( \widehat{m}^{(k)}-1\right) \geq 0 \\
%     %         &\sum_{i=1}^{k M_E-1}\frac{1}{\left(k M_E-i\right)} \hspace{10.2cm},\text{otherwise}
%     %         \end{cases}\\
%     %         \psi_{\infty}= 
% \psi_{\infty}(k)=
%     \begin{cases}
%         H_{kM_E-1}-\sum_{\mathbf{m} \in \mathcal{M}^{(k)}}\sum_{l=0}^{\widehat{m}^{(k)}-1}(-1)^{\widehat{m}^{(k)}-l-1}\binom{\widehat{m}^{(k)}-1}{l}\frac{\big(\prod_{i=1}^{k}{\Gamma\left(M_E+m_q\right)}\big)}{\left(kM_E+\widehat{m}^{(k)}-l-1\right)\big(\prod_{i=1}^{k}{m_q!}\big){\Gamma(M_E)}^k}&, \widehat{m}^{(k)} \neq 0 \\
%         H_{kM_E-1} &,\text{otherwise}
%         \end{cases}.
% \end{align}
% \begin{align}\label{eq_psi_OS_Asympt}
% \psi_{\infty}^{(k)}=
%     \begin{cases}
%         H_{kM_E-1}, & \widehat{m}^{(k)} = 0  \\
%         H_{kM_E-1}-I_1^{(k)}, & \widehat{m}^{(k)}\neq 0 
%         \end{cases}.
% \end{align}
\begin{align}\label{eq_psi_OS_Asympt}
\psi_{\infty}^{(k)}=
        H_{kM_E-1}-I_1^{(k)}.
\end{align}

From (\ref{eq_C_0_asympt_OS}) and (\ref{eq_Xi_first_time}), it is clear that the asymptotic ESR depends on the ratio $\lambda_D/\lambda_E$. This ratio arises in the expression for $I_0^{(k)}$; however, $I_1^{(k)}$ is independent of the ratio. This shows that one part of the ESR varies linearly with the ratio $\lambda_D/\lambda_E$ in the logarithmic scale, while the other part is independent of this ratio. As the ratio $\lambda_D/\lambda_E$ increases asymptotically, (\ref{eq_C_erg_L_1_OS_Asympt}) becomes independent of all system parameters except this ratio.
% conforms with the observation in the high-SNR approximation studied in the previous section. 
We now express the asymptotic ESR as a linear function of $\log_2(\lambda_D)$ as (c.f. \cite{Chinmoy_letters21,Lifeng_2013_C_erg_asymptotic}), 
% in order to clearly \textcolor{red}{show the dependence of the ESR on the system parameters as}
\begin{align}\label{C_erg_asymp_formula}
    &C_{\mathrm{erg}}^{\infty} = \mathcal{S}_\infty\left(\log_2\left (\lambda_{D}\right) - \mathcal{L}_\infty \right),
\end{align}
where $S_\infty$ is the slope, i.e., the maximum multiplexing gain or number of degrees of freedom, and $\mathcal{L}_\infty$ is the high-SNR offset. By observing \eqref{eq_C_erg_L_1_OS_Asympt}, the slope and high-SNR offset are given by 
\begin{align}\label{eq_C_erg_OS_asympt_slope_and_offset}
    % \mathcal{S}_\infty =\lim_{\lambda_{D}\rightarrow{\infty}}\frac{C_{erg}^{\infty}}{\log_2\left (\lambda_{D}\right)} = 1
     \mathcal{S}_\infty &=1,~\text{and}\nn\\
         % &\mathcal{L}_\infty = \lim_{\lambda_{D}\rightarrow{\infty}}\left({\log_2\left (\lambda_{D}\right)}-\frac{C_{erg}^{\infty}}{\mathcal{S}_\infty} \right) \nonumber\\
    %     & = \log_2\left (\lambda_E\right)-\frac{1}{\ln(2)} \sum_{k=1}^{K}(-1)^{k}\binom{K}{k} \left[\sum_{i=1}^{k M_E-1}\frac{1}{\left(k M_E-i\right)}+\psi_{\infty}\right].
   \mathcal{L}_\infty &=\log_2\left (\lambda_E\right)+\frac{1}{\ln(2)}\sum_{k=1}^{K}(-1)^{k+1}\binom{K}{k} \psi_{\infty}^{(k)},
\end{align}
respectively.

We conclude from (\ref{C_erg_asymp_formula}) and (\ref{eq_C_erg_OS_asympt_slope_and_offset}) in the high-SNR regime, the ESR varies linearly with $\lambda_D$ in the logarithmic  scale. The high-SNR slope is always unity and does not depend on $K$, $M_D$, or $M_E$. On the other hand, the high-SNR offset is independent of $\lambda_D$ and depends on $\lambda_E$, $M_E$, $M_D$, and $K$. We can also easily identify from (\ref{eq_C_erg_OS_asympt_slope_and_offset}) that as $\lambda_E$ increases, the high-SNR offset increases, and as a result, the ESR decreases.

When $K>1$ and $M_D=M_E$, we see from (\ref{eq_psi_OS_Asympt}) that the high-SNR offset varies with $M_D$.
% \textcolor{blue}{Also for $M_D=M_E=X$, $\mathcal{L}_\infty$ is lesser than that for $M_D=M_E>X$. As a result the ESR performance is better in former case.} 
This means that the number of multipath components has a significant effect on the ESR performance of the OS scheme. 
However, as a reference, we note that in the special case when $K=1$ and $M_D=M_E$, we have $\psi_{\infty}^{(1)}=0$, since 
\begin{align}\label{eq_psi_OS_Asympt_K=1}
   \psi_{\infty}^{(1)}=
       H_{M_E-1}-H_{M_D-1},
\end{align}
due to $I_1^{(1)}=\sum_{m_1=1}^{M_D-1} \frac{1}{m_1}=H_{M_D-1}$ in \eqref{eq_Xi_first_time}. 
% when $K=1$, $I_1^{(k)}$ in \eqref{eq_Xi_first_time} becomes $I_1^{(1)}=\sum_{m_1=1}^{M_D-1} \frac{1}{m_1}=H_{M_D-1}$. Thus, (\ref{eq_psi_OS_Asympt}) is expressed as % \begin{align}\label{eq_psi_OS_Asympt_K=1}
% %   \psi_{\infty}^{(1)}=
% %     \begin{cases}
% %       H_{M_E-1}, & \widehat{m}^{(k)} = 0  \\
% %          H_{M_E-1}-H_{M_D-1}, &    \widehat{m}^{(k)} \ne 0 
% %         \end{cases}.
% % \end{align}
% \begin{align}\label{eq_psi_OS_Asympt_K=1}
%   \psi_{\infty}^{(1)}=
%       H_{M_E-1}-H_{M_D-1}.
% \end{align}
% Now if $M_D=M_E$,  $\psi_{\infty}^{(1)}=0$ and
Hence, the offset  $\mathcal{L}_\infty =\log_2\left (\lambda_E\right)$ is the same as the offset under Rayleigh fading channel conditions, i.e., $M_D=M_E=1$. This means for a single-transmitter system with an arbitrary but equal number of multipath components in the destination and eavesdropper channels, the ESR will asymptotically approach the narrowband Rayleigh fading channel ESR.
% and will depend only on the destination and eavesdropper channel average SNR per multipath component. 
% However, when $K>1$ and $M_D=M_E$, we see from (\ref{eq_psi_OS_Asympt}) that the high-SNR offset varies with $M_D$. This means that the number of multipath components has a significant effect on the ESR performance of the OS scheme.

% of $C_{erg}^\infty$ will depend only on the destination and eavesdropper SNR per multipath component. Conversely when $K \neq 1$, the slope will depend on all other factors except $\lambda_D$.

For the case of Rayleigh fading channels, i.e., $M_D=M_E=1$, with arbitrary $K$, from \eqref{eq_psi_OS_Asympt} we have $\psi_{\infty}^{(k)}=H_{k-1}$ as $I_1^{(k)}$ exists only for $\widehat{m}^{(k)} \neq 0$. Hence, from \eqref{eq_C_erg_OS_asympt_slope_and_offset}, the slope is $\mathcal{S}_\infty = 1$ and the offset is \begin{align}\label{eq_C_erg_OS_asympt_slope_and_offset_any_K_M_D=M_E=1}
    \mathcal{L}_\infty &=\log_2 (\lambda_E)
    -\frac{1}{\ln(2)}\sum_{k=1}^{K}(-1)^{k}\binom{K}{k} H_{k-1}
    \nn\\
    & 
    =\log_2 (\lambda_E)
    -\frac{1}{\ln(2)}H_{K-1}.
    \end{align}
This special case expression for the ESR exactly matches the ESR expression derived in \cite{Chinmoy_letters21} for the OS scheme with unreliable backhaul and multiple eavesdroppers in Rayleigh fading when the backhaul is considered to be perfect, and there is a single eavesdropper.

\section{ESR of Sub-optimal Source-Destination Pair Selection }\label{sec_SS}
The SS scheme is implemented when the CSI of the link $S_k$-$E$ is not available for any $k \in \mathcal{S}$. In this scenario, the source-destination pair is selected for which $\gamma_{D}^{(k,l)}$ is maximum among all $k$ and $l$. In this scheme, $\Gamma_{S}$ is written as 
 \begin{align}\label{eq_Gamma_S_SS}
\Gamma_{S}=\frac{1+\max\limits_{\forall{k},\forall{l}}\big\{\gamma_{D}^{(k,l)}\big\}}{1+\gamma_E^{(k^*)}},
\end{align} 
where $\gamma_E^{(k^*)}$ is the SNR at $E$ corresponding to the selected transmitter $k^*$. As $S_k$-$E$ for each $k$ is i.i.d., $\gamma_E^{(k^*)}$ follows the distribution expressed in (\ref{CDF_S_D}). In Section \ref{sec_OS}, we proposed a way to evaluate the ESR of the OS scheme; in this section, we will show that it can also be extended for the SS scheme. 
We first proceed with the evaluation of the CDF of $\Gamma_{S}$.
%\vspace{-0.1cm}
\subsection{Determining the CDF of  $\Gamma_{S}$}
As $\Gamma_{S}^{(k,l)}$ for all $k \in \mathcal{S}$ and $l \in \mathcal{D}$ is i.i.d., the CDF of $\Gamma_{S}$ is derived as 
\begin{align}\label{eq_CDF_SS_max_SNR_at_D}
 F_{\Gamma_{S}}(x)
%  =\mathbb P \{\Gamma_S \leq x\}
&=\mathbb P\lb[\max\limits_{\forall{k},\forall{l}}\big\{\gamma_{D}^{(k,l)}\big\}\le x\left({1+\gamma_E^{(k^*)}}\right)-1\rb]\nn\\
 &= \int_0^\infty\left[F_{\gamma_{D}^{(k,l)}}\left(x\left({1+y}\right)-1\right)\right]^{KL}
f_{\gamma_E^{(k^*)}}\left(y \right)dy\nn\\
&= 1-\sum_{k=1}^{KL}(-1)^{k+1}\binom{KL}{k}\Psi^k(x),
\end{align}
where 
\begin{align}\label{eq_Psi_x_k_SS}
    \Psi^k(x)&=\int_0^\infty\Psi^k(x,y)f_{\gamma_E^{(k^*)}}(y)dy, 
\end{align}
and
\begin{align}\label{eq_Psi_SS}
&\Psi^k(x,y)\nn\\
&=\Bigg[\sum_{m=0}^{M_D-1}\sum_{n=0}^{m}\binom{m}{n}\frac{\left(x-1\right)^{m-n}x^{n}y^{n}\exp\big(-{\frac{x\left(1+y \right)-1}{\lambda_{D}}}\big)}{(m!){\lambda_D}^{m}}\Bigg]^k\nn\\
    &=\sum_{\mathbf{m} \in \mathcal{M}^{(k)}}\sum_{\mathbf{n} \in \mathcal{N}^{(k)}(\mathbf{m})}\sum_{j=0}^{\widehat{m}^{(k)}-\widehat{n}^{(k)}}(-1)^{\widehat{m}^{(k)}-\widehat{n}^{(k)}-j}
    \nn\\
    &\times\binom{\widehat{m}^{(k)}-\widehat{n}^{(k)}}{j}
    \frac{\Big(\prod\limits_{q=1}^{k}\binom{m_{q}}{n_q}\Big)x^{\widehat{n}^{(k)}+j}\exp\big(-{\frac{k(x-1)}{\lambda_{D}}}\big)}{\Big(\prod\limits_{q=1}^{k}m_{q}!\Big){\lambda_D}^{\widehat{m}^{(k)}}}\nn\\
    &\times y^{\widehat{n}^{(k)}}\exp\big(-{\frac{kxy}{\lambda_{D}}}\big).
\end{align}
The derivation of $\Psi^k(x,y)$ in \eqref{eq_Psi_SS} using \eqref{eq_cross_mul_3_sums} is similar to the method adopted in the OS scheme. Then, the solution of the integral in \eqref{eq_Psi_x_k_SS} is obtained with the help of \cite[eq. (3.351.3)]{ryzhik_2007} as
\begin{align}\label{eq_psi_k_SS_Expanded}
\Psi^k(x) 
% =&\int_0^\infty\Psi^k(x,y)f_{\gamma_E^{(k)}}(y)dy\nn\\
&=\sum_{\mathbf{m} \in \mathcal{M}^{(k)}}\sum_{\mathbf{n} \in \mathcal{N}^{(k)}(\mathbf{m})}\sum_{j=0}^{\widehat{m}^{(k)}-\widehat{n}^{(k)}}(-1)^{\widehat{m}^{(k)}-\widehat{n}^{(k)}-j}\nonumber\\
    &\times \binom{\widehat{m}^{(k)}-\widehat{n}^{(k)}}{j} \frac{{\lambda_D}^{M_E-(\widehat{m}^{(k)}-\widehat{n}^{(k)})}\Gamma(M_E+\widehat{n}^{(k)} )}{{\lambda_E}^{M_E}k^{M_E+\widehat{n}^{(k)}}\Big(\prod\limits_{q=1}^{k}{m_{q}!}\Big)\Gamma(M_E)}\nn\\
     &\times  \frac{\Big(\prod\limits_{q=1}^{k}\binom{m_{q}}{n_q}\Big)x^{\widehat{n}^{(k)}+j}\exp\big(-\frac{k(x-1)}{\lambda_D}\big)}{\big(x+\frac{\lambda_D}{k\lambda_E}\big)^{M_E+\widehat{n}^{(k)}}}.
\end{align}

\subsection{Evaluation of ESR}
Substituting \eqref{eq_psi_k_SS_Expanded} into  \eqref{eq_ESR_CDF_eqn}, the ESR in SS can be expressed as
\begin{align}\label{eq_ESR_SS_int}
    C_{erg}&=\frac{1}{\ln(2)}\sum_{k=1}^{KL}(-1)^{k+1}\binom{KL}{k}\int_1^\infty\frac{\Psi^k(x)}{x}dx\nn\\
    &=\frac{1}{\ln(2)}\sum_{k=1}^{KL}(-1)^{k+1}\binom{KL}{k}\big[I_0^{(k)}+I_1^{(k)}\big],
\end{align}
where $I_0$ corresponds to the integral with condition  $\widehat{n}^{(k)}+j=0$ and is expressed as 
\begin{align}\label{eq_I_0_SS_int}
I_0^{(k)}&=\sum_{\mathbf{m} \in \mathcal{M}^{(k)}} (-1)^{\widehat{m}^{(k)}}\frac{{\lambda_D}^{M_E}}{{\lambda_{D}}^{\widehat{m}^{(k)}}{k}^{M_E}{\lambda_E}^{M_E}\Big(\prod\limits_{q=1}^{k}{m_{q}!}\Big)}\nn\\
&\times \int_{1}^{\infty}\frac{\exp\big(-\frac{k(x-1)}{\lambda_D}\big)}{x\big(x+\frac{\lambda_{D}}{k\lambda_{E}}\big)^{M_E}}dx,
\end{align}
and $I_1$ correspond to the integral with condition $\widehat{n}^{(k)}+j > 0$  and is expressed as
\begin{align}\label{eq_I_1_SS_int}
&I_1^{(k)}=\sum_{\mathbf{m} \in \mathcal{M}^{(k)}}\sum_{\mathbf{n} \in \mathcal{N}^{(k)}(\mathbf{m})}\sum_{j=0}^{\widehat{m}^{(k)}-\widehat{n}^{(k)}}(-1)^{\widehat{m}^{(k)}-\widehat{n}^{(k)}-j}\nonumber\\
    &\times\binom{\widehat{m}^{(k)}-\widehat{n}^{(k)}}{j} \frac{{\lambda_D}^{M_E-(\widehat{m}^{(k)}-\widehat{n}^{(k)})}\Gamma(M_E+\widehat{n}^{(k)} )}{{\lambda_E}^{M_E} k^{M_E+\widehat{n}^{(k)}}\Big(\prod\limits_{q=1}^{k}{m_{q}!}\Big)\Gamma(M_E)}\nn\\
    &\times \int_{1}^{\infty}\frac{\Big(\prod\limits_{q=1}^{k}\binom{m_{q}}{n_q}\Big)x^{\widehat{n}^{(k)}+j-1}\exp\big(-\frac{k(x-1)}{\lambda_D}\big)}{\big(x+\frac{\lambda_D}{k\lambda_E}\big)^{M_E+\widehat{n}^{(k)}}}dx.
\end{align}
The solutions of $I_0^{(k)}$ and $I_1^{(k)}$ are provided into \eqref{eq_I_0_SS} and \eqref{eq_I_1_SS} respectively. With the substitution of \eqref{eq_I_0_SS} and \eqref{eq_I_1_SS} in \eqref{eq_ESR_SS_int}, the ESR is expressed in closed form. Here we note that in (\ref{eq_ESR_SS_int}) if the total number of links $KL$  is the same for any combination of $K$ and $L$, the ESR performance would be the same for these combinations. Thus increasing either $K$ or $L$ is identical to the system secrecy in the SS scheme.

From (\ref{eq_ESR_SS_int}) it is difficult to gain insights. Hence, we provide a high-SNR approximation and asymptotic analysis in the following sections, as was provided for the OS scheme in Sections \ref{section_High_SNR_Approximation} and \ref{subsection_Asymptotic_Analysis_of_ESR_in_OS}, respectively.

\begin{table*}
\begin{align}\label{eq_I_0_SS}
    I_0^{(k)}&= \sum_{\mathbf{m} \in \mathcal{M}^{(k)}} (-1)^{\widehat{m}^{(k)}}\frac{{\lambda_D}^{M_E-\widehat{m}^{(k)}}\exp\Big({\frac{k}{\lambda_{D}}}\Big)}{{\lambda_E}^{M_E}{k}^{M_E}\Big(\prod\limits_{q=1}^{k}{m_{q}!}\Big)}\nn\\
    &\times \bigg[\Gamma\Big(0,\frac{k }{\lambda_D} \Big)-\exp\Big({\frac{1}{\lambda_{E}}}\Big)\sum_{t=1}^{M_E}{{\lambda_E}^{t-M_E}}\Gamma\Big(-\left(M_E-t\right),\frac{k }{\lambda_D}\Big(1+\frac{\lambda_D}{k \lambda_E}\Big)\Big)\bigg].\\
% \end{align}
% \hrule
% \end{table*}
% \begin{table*}
% \begin{align}
\label{eq_I_1_SS}
    I_1^{(k)}&=\sum_{\mathbf{m} \in \mathcal{M}^{(k)}} \sum_{\mathbf{n} \in \mathcal{N}^{(k)}(\mathbf{m})}\sum_{v=0}^{\widehat{m}^{(k)}-\widehat{n}_q^{(k)}}\sum_{i=0}^{\widehat{n}_q^{(k)}+v-1}(-1)^{\widehat{m}^{(k)}+\widehat{n}_q^{(k)}+v+i}\binom{\widehat{m}^{(k)}-\widehat{n}_q^{(k)}}{v}\binom{\widehat{n}_q^{(k)}+v-1}{i}\nonumber\\
   &\times\frac{\Gamma(M_E+\widehat{n}_q^{(k)})\Big(\prod\limits_{q=1}^{k}\binom{m_{q}}{{n}_q}\Big)\Gamma\Big(-\big(k M_E-v+i\big),\frac{k}{\lambda_D}\big( 1+\frac{\lambda_D}{k\lambda_E}\big) \Big)\exp\big(\frac{k}{\lambda_D}+\frac{1}{\lambda_E}\big)}{{\lambda_D}^{\widehat{m}^{(k)}-\widehat{n}_q^{(k)}-v}{\lambda_E}^{M_E+i}k^{\widehat{n}_q^{(k)}+v}\Big(\prod\limits_{q=1}^{k}{m_{q}!}\Big)\Gamma(M_E)}.
\end{align}
\hrule
\end{table*}

\subsection{High-SNR Approximation}\label{Subsection:ESR High SNR Analysis}
In the high-SNR regime, as in the OS scheme,  we assume  $\max\limits_{\forall{k},\forall{l}}\{\gamma_{D}^{(k,l)}\}>>1$ and $\gamma_{E}^{(k^*)}>>1$ for all $k$ and $l$, thus we use the approximation $\Gamma_S^{(k,l)}\approx\frac{\max\limits_{\forall{k},\forall{l}}\{\gamma_{D}^{(k,l)}\}}{\gamma_{E}^{(k^*)}}$ by neglecting unity from the numerator and denominator in \eqref{eq_Gamma_S_SS}. As discussed in the OS scheme, this forms an upper bound on the exact $\Gamma_S^{(k,l)}$. 
With this approximation, the high-SNR CDF of $\Gamma_{S}$ is evaluated as
\begin{align}\label{eq_CDF_Gamma_S_SS_High_SNR}
F_{\Gamma_{S}}(x) &=\mathbb P\lb[\max\limits_{\forall{k},\forall{l}}\big\{\gamma_{D}^{(k,l)}\big\}\le x\left({\gamma_E^{(k^*)}}\right)\rb]\nn\\
&=\int_0^\infty \lb[F_{\gamma_{D}^{(k,l)}} \left(x y\right)\rb]^{KL} f_{\gamma_E^{(k^*)}}(y) dy.
% \nn\\
%   &= 1-\int_0^\infty\sum_{k=1}^{KL}(-1)^{k+1}\binom{KL}{k} \Psi^k(x,y)f_{\gamma_E^{(k^*)}}\left(y \right)dy\nn\\
%   &=1-\sum_{k=1}^{KL}(-1)^{k+1}\binom{KL}{k}\Psi^k(x),
\end{align}
This integral can be solved exactly as in \eqref{eq_CDF_SS_max_SNR_at_D} where $\Psi^k(x)$ 
 has the same form as in \eqref{eq_Psi_x_k_SS}. Here, $\Psi^k(x,y)$ as in \eqref{eq_Psi_SS} can be written as 
 \begin{align}\label{eq_Psi_SS_High_SNR}
\Psi^k(x,y)&=\bigg[\sum_{m=0}^{M_D-1}\frac{x^{m} y^{m} \exp\big(-\frac{xy}{\lambda_D}\big)}{{(m!)\lambda_D}^{m}}\bigg]^k 
\nn\\ 
&
=\sum_{\mathbf{m} \in \mathcal{M}^{(k)}} \frac{x^{\widehat{m}^{(k)}} y^{\widehat{m}^{(k)}} \exp\big(-\frac{kxy}{\lambda_D}\big)}{\Big(\prod\limits_{q=1}^{k}{m_{q}!}\Big){\lambda_D}^{\widehat{m}^{(k)}}}.
\end{align}
 With the help of \eqref{eq_Psi_SS_High_SNR}, $\Psi^k(x)$ is evaluated in a similar manner to \eqref{eq_psi_k_SS_Expanded} as 
\begin{align}\label{eq_psi_k_high_SNR_SS_expanded}
    \Psi^k(x)&=\sum_{\mathbf{m} \in \mathcal{M}^{(k)}}\frac{{\lambda_D}^{M_E}{\Gamma\left(M_E+\widehat{m}^{(k)}\right)} }{{\lambda_E}^{M_E}k^{{M_E}+\widehat{m}^{(k)}}\Big(\prod\limits_{q=1}^{k}m_{q}!\Big){\Gamma\left(M_E\right)}}\nn\\
    & \times \frac{x^{\widehat{m}^{(k)}}}{\big(x+\frac{\lambda_D}{k \lambda_E}\big)^{M_E+\widehat{m}^{(k)}}}.
\end{align}
In the above expression, the product-of-sums is converted to the sum-of-products for integration using an approach similar to \eqref{eq_cross_mul_3_sums} as in the OS scheme. 
% The solution of the integral in  \eqref{eq_CDF_Gamma_S_SS_High_SNR} provides the CDF of $\Gamma_S^{(k,l)}$ in the form of \eqref{eq_CDF_Gamma_S_SS_High_SNR},
% \begin{align}\label{eq_CDF_SS_Psi_High_SNR}
%   &F_{\Gamma_{S}}(x)=1-\sum_{k=1}^{K}(-1)^{k+1}\binom{K}{k}\Psi^k(x),
% \end{align}
Next, the ESR can be expressed in the form of \eqref{eq_ESR_SS_int}. In \eqref{eq_ESR_SS_int} for high-SNR approximation, $I_0^{(k)}$ corresponds to the condition $\widehat{m}^{(k)} = 0$, with solution given in \eqref{eq_I_0_SS_High_SNR}, while $I_1^{(k)}(x)$ corresponds to all other cases,  i.e., $\widehat{m}^{(k)} \neq 0$, the solution of which is presented in \eqref{eq_I_1_SS_High_SNR}.
\begin{table*}
\begin{align}\label{eq_I_0_SS_High_SNR}
    I_0^{(k)}&= {\ln \big(1+ \frac{\lambda_D}{k \lambda_E}\big)-\sum_{t=1}^{M_E-1}}\frac{\big(\frac{\lambda_D}{k\lambda_E}\big)^{M_E-t}}{\left(M_E-t\right)\big(1+\frac{\lambda_D}{k\lambda_E} \big)^{M_E-t}}.\\
% \end{align}
% \hrule
% \end{table*}
% \begin{table*}
%\vspace{-0.2cm}
% \begin{align}
\label{eq_I_1_SS_High_SNR}
    I_1^{(k)}&=\sum_{\mathbf{m} \in \mathcal{M}^{(k)}}\sum_{v=0}^{\widehat{m}^{(k)}-1}(-1)^{\widehat{m}^{(k)}-v-1}\binom{\widehat{m}^{(k)}-1}{v}\nn\\
    &\times \frac{\big(\frac{\lambda_D}{k\lambda_E}\big)^{(M_E+\widehat{m}^{(k)}-v-1)}{\Gamma(M_E+\widehat{m}^{(k)})}}{\big(1+\frac{\lambda_D}{k\lambda_E} \big)^{(M_E+\widehat{m}^{(k)}-v-1)}k^{\widehat{m}^{(k)}}(M_E+\widehat{m}^{(k)}-v-1)\Big(\prod\limits_{q=1}^{k}m_{q}! \Big)\Gamma(M_E)}.
\end{align}
\hrule
\end{table*}

% Comparing  the solution of (\ref{eq_ESR_OS_High_SNR_Psi_by_x}) for the OS scheme with the solution of \eqref{eq_ESR_SS_int} for the SS scheme, respectively, we note that the ESR is a function of $\lambda_D/\lambda_E$ in both selection schemes. 

% Comparing the high-SNR single destination case of OS scheme in \eqref{eq_ESR_OS_High_SNR_Psi_by_x} and SS scheme in \eqref{eq_ESR_SS_int} we observe that in contrast to OS, the SNR ratio in the SS scheme is lower because of the scaling factor $k$. This intuitively relates to the fact that the SS scheme achieves a lower average ratio of destination channel SNR to Eavesdropper channel SNR per path than the OS scheme, and therefore its ESR performance suffers.

Comparing  (\ref{eq_I_0_L_1_OS_High_SNR}) and (\ref{eq_I_1_L_1_OS_High_SNR}) for the OS scheme with \eqref{eq_I_0_SS_High_SNR} and \eqref{eq_I_1_SS_High_SNR} for the SS scheme for $L=1$, respectively, we note that the ESR is a function of $\lambda_D/\lambda_E$ in both selection schemes. However, we observe that the SNR ratio is lower in the SS scheme as it is decreased by the factor $k$ number of transmitters. This intuitively relates to the fact that the SS scheme achieves a lower average destination to eavesdropper channel SNR ratio per path than the OS scheme, and therefore its ESR performance suffers.

\subsection{Asymptotic Analysis}\label{subsection_Asymptotic_Analysis_of_ESR_in_SS}
% With the same assumptions of $\lambda_D \rightarrow{\infty}$ for a given $\lambda_E$
As in Section \ref{subsection_Asymptotic_Analysis_of_ESR_in_OS}, the asymptotic analysis in SS is aimed at providing better insights regarding the relationship between the system parameters and the ESR. By approximating $1+\frac{\lambda_{D}}{k\lambda_{E}} \approx \frac{\lambda_{D}}{k\lambda_{E}}$ in \eqref{eq_I_0_SS_High_SNR} and (\ref{eq_I_1_SS_High_SNR}) for all $k\in\{1 ,\ldots, K\}$, $I_0^{(k)}$ for the SS scheme in the asymptotic scenario is evaluated as 
% \begin{table*}
\begin{align}\label{eq_I_0_asympt_SS}
    I_0^{(k)}&=\ln\Big(\frac{\lambda_D}{k\lambda_E}\Big)-H_{M_E-1}.
\end{align}
 Similarly, after performing some mathematical simplifications, we obtain $I_1^{(k)}$ as
\begin{align}
\label{eq_I_1_asympt_SS}
    I_1^{(k)}&=
    \sum_{\mathbf{m} \in \mathcal{M}^{(k)}}\frac{\Gamma(\widehat{m}^{(k)})}{k^{\widehat{m}^{(k)}}\Big(\prod\limits_{q=1}^{k}{m_{q}!}\Big)}.  
\end{align}
We also transform $C_{erg}^{\infty}$ into a linear function of $\log_2(\lambda_D)$ in the logarithmic scale as was done in (\ref{C_erg_asymp_formula}) for the OS scheme.
The slope $\mathcal{S}_\infty$ and  the offset $\mathcal{L}_\infty$ for the SS scheme are evaluated as
\begin{align}\label{eq_C_erg_SS_asympt_slope_and_offset}
 \mathcal{S}_\infty&=1, ~~\text{and}~~
 \nn\\
    \mathcal{L}_\infty& =\sum_{k=1}^{KL}(-1)^{k+1}\binom{KL}{k}\Big[\log_2(k\lambda_E)+\frac{1}{\ln(2)}\psi_{\infty}^{(k)}\Big],
\end{align}
where $\psi_{\infty}^{(k)}$ is expressed as
\begin{align}\label{eq_psi_SS_Asympt}
   \psi_{\infty}^{(k)}=H_{M_E-1}-I_1^{(k)}. 
\end{align}

% We observe from the asymptotic analysis of the SS scheme that the slope is independent of all system parameters and is equal to unity, as in the OS scheme for $L=1$ in (\ref{eq_C_erg_OS_asympt_slope_and_offset}). The high-SNR offset depends on $\lambda_E$, $M_E$, $M_D$, $K$ and $L$.
% % Further, it is visible by comparing the results in (\ref{eq_C_erg_L_1_OS_Asympt}) and (\ref{eq_I_0_asympt_SS}) that the SS scheme achieves a lower average ratio of destination channel SNR to eavesdropper channel SNR per path than the OS scheme, as it is attenuated by a factor of $k$; this was also observed in the high-SNR approximate analysis.
% Further, it is clearly visible by comparing the results in (\ref{eq_C_erg_L_1_OS_Asympt}) and (\ref{eq_I_0_asympt_SS})
% that the SS scheme achieves lower average destination to eavesdropper channel SNR per path
% ratio than the OS scheme, as was observed also in the high-SNR approximate analysis.

We observe from the asymptotic analysis of the SS scheme that the slope is independent of all system parameters and is equal to unity, as in the OS scheme for $L=1$ in (\ref{eq_C_erg_OS_asympt_slope_and_offset}). The high-SNR offset depends on $\lambda_E$, $M_E$, $M_D$ and $K$. Further, it is clearly visible by comparing the results in (\ref{eq_I_0_asympt_SS}) and (\ref{eq_C_0_asympt_OS}) that the SS scheme achieves lower average destination to eavesdropper channel SNR per path ratio than the OS scheme, as was also observed in the high-SNR approximate analysis. 

From \eqref{eq_psi_SS_Asympt} it is  noted that even if $M_D=M_E$ when $K,L > 1$, even the high-SNR offset is different for a different number of multipath components. This is the same as observed in the OS scheme for $L=1$, even though the values are different from those for the OS scheme. In the special case when $K=1$, the OS and SS scheme results are the same as there is no selection involved. % $\psi_{\infty}^{(1)}$ is the same as in \eqref{eq_psi_OS_Asympt_K=1} as in the OS scheme. In this case when $M_D=M_E$, $\psi_{\infty}^{(1)}=0$ and the offset is $\log_2(\lambda_E)$. Thus, $C_{erg}^{\infty}$ only depends on $\lambda_D/\lambda_E$ as in the OS scheme.
For the case of Rayleigh fading, i.e., $M_D=M_E=1$ with an arbitrary $K$, $\psi_{\infty}^{(k)}=0$ as $H_{0}=0$ and $I_1^{(k)}=0$ in \eqref{eq_C_erg_SS_asympt_slope_and_offset} ($I_1^{(k)}$ exists only for the condition $\widehat{m}^{(k)} \neq 0$). This substitution in \eqref{eq_C_erg_SS_asympt_slope_and_offset} provides the slope as $\mathcal{S}_\infty=1$ and the high-SNR offset as
\begin{align}\label{eq_C_erg_SS_asympt_slope_and_offset_any_K_M_D=M_E=1}
 \mathcal{L}_\infty  &=\sum_{k=1}^{KL}(-1)^{k+1}\binom{KL}{k}\log_2(k\lambda_E)\nn\\
%  \nn\\
%  \mathcal{L}_\infty
&=\log_2(\lambda_E)-\frac{1}{\ln(2)}\sum_{k=1}^{KL}(-1)^{k}\binom{KL}{k}\ln(k),
 \end{align}
 for the SS scheme.
On comparing \eqref{eq_C_erg_SS_asympt_slope_and_offset_any_K_M_D=M_E=1} and  \eqref{eq_C_erg_OS_asympt_slope_and_offset_any_K_M_D=M_E=1}, it can be noted that since $H_{k-1}>\ln(k)$ for all $k\in \mathcal{S}$, the high-SNR offset in the OS scheme is greater than that of the SS scheme in the case of Rayleigh fading. This leads to improved performance in the OS scheme. 

As on the application of SC-CP modulation, the underlying frequency selective fading channel behaves  as a  narrowband  Nakagami-$m$ fading  channel  with  integer parameter $m$, the ESR results derived for the OS and SS schemes in this paper also provide the corresponding analysis for the case of narrowband channels with Nakagami-$m$ fading channels. In this regard, we observe that \cite{Mallik_TWC_14} considers the sub-optimal transmitter selection scheme over Nakagami-$m$ fading channel with multiple antennas at the destination and eavesdropper using diversity combining. However, the source-destination pair selection is not present in \cite{Mallik_TWC_14}. 
% , our ESR result matches with \cite{Mallik_TWC_14} when all receivers have a single antenna. 
Moreover, the methodology in \cite{Mallik_TWC_14} cannot be extended to analyze the OS scheme.  
% The complexity of the ESR is also not discussed. In addition, the asymptotic analysis uses approximations for the effective source to destination  SNR, whereas we consider the exact distributions. Moreover, our derivation methodology is general and can be applied to a wide variety of transmitter selection schemes. 

\section{Transmitter and Path Correlation}\label{sec_Spatial_Correlation}

% The effects of transmitter and path correlation on transmitter-destination pair selection in frequency selective fading channels remain to be explored. With this motivation, we present the ESR analysis through simulation only.

In this section, we will describe the transmitter and path correlation models that will be adopted in our work. In Section \ref{sec_Results} we will use these correlation models together with our derived analytical results to study the effect of these correlations on the ESR performance. We will consider transmitter correlation, destination path correlation, and eavesdropper path correlation. The composite destination channel $\mathbf{H}_{D}$ of size $L\times KM_D$, including all multipath components $i\in\{1 ,\ldots, M_D\}$ between $S_k$, for all $k\in \mathcal{S}$, and $D_l$, for all $l\in\mathcal{D}$, is generated following \cite{Space_tap_Corr_TWC_2007} by introducing transmitter and path correlation as
\begin{align}\label{eq_H_D_total_2}    
\mathbf{H}_{D}=\mathbf{H}_{\omega(L \times K M_D)}\left(\mathbf{R}_{D}^{\text{T}/2} \otimes \mathbf{R}_{S}^{1/2}\right),
\end{align}
where $\mathbf{H}_{\omega{(L \times K M_D)}}$ denotes the $L \times K M_D$ channel matrix with each element $h_{\omega}^{(k,l)}(i)$ is a circularly symmetric complex Gaussian random variable with zero mean and variance $\lambda_D$ for all possible $k\in\{1 ,\ldots, K\}$, $l\in\mathcal{D}$, and $i\in\{1 ,\ldots, M_D\}$. $\mathbf{H}_{\omega{(L \times K M_D)}}$ is denoted as 
\begin{align}\label{eq_H_i_def}    
\mathbf{H}_{\omega(L\times KM_D)}&=
\begin{bmatrix} 
\mathbf{h}^{(1,1)}_{\omega}& &\ldots&  &\mathbf{h}^{(1,L)}_{\omega}\\
\vdots&  &\mathbf{h}^{(k,l)}_{\omega}&  &\vdots\\
\mathbf{h}^{(K,1)}_{\omega}&  &\ldots&  &\mathbf{h}^{(K,L)}_{\omega}
  \end{bmatrix}^\text{T},
\end{align}
where $\mathbf{h}_{\omega}^{(k,l)}=\left[h_{\omega}^{(k,l)}(1),\ldots, h_{\omega}^{(k,l)}(M_D)\right]^\text{T}$.
% \begin{align}\label{eq_h_omega_def} 
%  \mathbf{h}_{\omega}^{(k,l)}=\left[h_{\omega}^{(k,l)}(1),\ldots, h_{\omega}^{(k,l)}(M_D)\right]^\text{T}
% \end{align}
$\mathbf{R}_S$ and $\mathbf{R}_D$ are transmitter correlation and destination path correlation matrices of size $K \times K$ and $M_D \times M_D$, respectively. These are Toeplitz matrices
with $(i,j)$-entry equal to
\begin{align}
\label{eq_R_D_def}
\left[\mathbf{R}_D\right]_{i,j}&=\lambda_{D}~\rho_{D}^{|i-j|};~\forall i,j \in\{1,\ldots, M_D\},\nn\\
% \label{eq_R_S_def}
\left[\mathbf{R}_S\right]_{i,j}&=\rho_{S}^{|i-j|};~\forall i,j \in\{1,\ldots, K\},
\end{align}
% where $\left[R_D\right]_{i_1,i_2}$ denotes the element of $R_D$ in the $i_1^{\text{th}}$ row and $i_2^{\text{th}}$ column, and
% \begin{align}\label{eq_R_D_def}
% \rho_{D_{i_1,i_2}}=\rho_D^{|i_1-i_2|},
% \end{align}
respectively, where $\rho_D$ and $\rho_S$ are correlation exponents with $0\leq|\rho_D|\leq 1$ and $0\leq|\rho_S|\leq 1$ \cite{Space_tap_Corr_TWC_2007}. In this scenario, 
% $\mathbf{H}_D$ is \begin{align}\label{eq_H_D_total_1}
$\mathbf{H}_{D}=\left[\mathbf{H}_1,\ldots, \mathbf{H}_{M_{D}}\right]$
% \end{align}
where $\mathbf{H}_i\in\mathbb{C}^{L \times K}$ denotes the channel matrix of the $i^{\text{th}}$  path of $S_k$-$D_l$ for all $k$ and $l$.  The correlation matrix between two different paths $i$ and $j$ $(i \neq j)$ can be defined as
\begin{align}
    &\mathbb{E}\left[\text{vec}\{\mathbf{H}_{i}\}~ \text{vec}\{\mathbf{H}_{j}\}^{\text{H}} \right]\nn\\
    &= \lambda_D ~ \rho_{D}^{|i-j|}~ \mathbf{R}_{S}^{\text{T}}; ~~\forall i,j \in\{1,\ldots, M_D\}.
\end{align}
$\mathbf{H}_{D}$ is structurally similar to (\ref{eq_H_i_def}) and we obtain $\mathbf{H}_{D}$ by replacing $\omega$ in $\mathbf{H}_{\omega}$ with $D$. Similarly, the composite eavesdropping channel $\mathbf{H}_{E}$ of size $1\times KM_E$, including all multipath components $i\in\{1 ,\ldots, M_E\}$ between $S_k$ and $E$ for all $k\in \mathcal{S}$, is generated including transmitter and path correlations assuming the eavesdropping path correlation matrix to be a Toeplitz matrix 
% \begin{align}\label{eq_R_E_def}
$\left[\mathbf{R}_E\right]_{i,j}=\rho_{E}^{|i-j|},~\forall i,j \in\{1,\ldots, M_E\}$,
% \end{align}
where  $0\leq|\rho_E|\leq 1$, by suitably modifying (\ref{eq_H_D_total_2})-(\ref{eq_R_D_def}).

Using channel realizations realizations based on $\mathbf{H}_{D}$  and $\mathbf{H}_{E}$ for all multipath components  between $S_k$ and $D_l$, and between $S_k$ and $E$, respectively, for all $k\in \mathcal{S}$ and $l\in\mathcal{D}$,  the simulation of ESR is performed by evaluating $\Gamma_S$  in \eqref{eq_Gamma_s} for the OS scheme and \eqref{eq_Gamma_S_SS} for the SS schemes and using \eqref{eq_ESR_basic_eqn}.

\section{Results}\label{sec_Results}
In this section, we provide analytical results along with the corresponding simulations except for Fig. \ref{CORR_graph} where only simulation results are plotted. Fig. \ref{fig_ESR_vs_SNR_variation_in_K_and_L_MD_ME_4} and \ref{fig_ESR_vs_M_D_SS_OS_fixed_K=L=2} do not include transmitter or path correlation, whereas Fig. \ref{CORR_graph} includes both transmitter correlation and path correlation. The simulation results completely agree with the analysis in all these figures, thus validating the correctness of our analytical approach.

\begin{figure}
 \centering
    \includegraphics[width=3in]{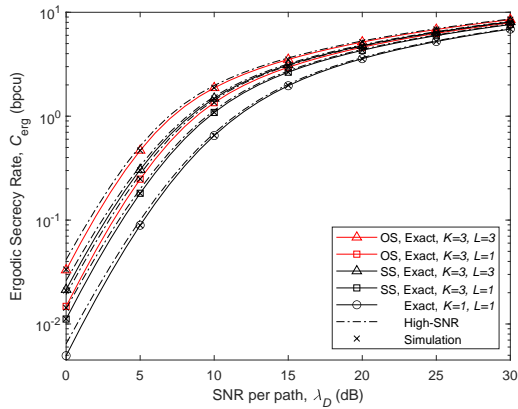}
 \caption{Variation of the exact ESR and its high-SNR approximation with $\lambda_D$ for $M_D=M_E=3$ and $\lambda_E=9$ dB.}
 \vspace{-0.4cm}
 \label{fig_ESR_vs_SNR_variation_in_K_and_L_MD_ME_4}
 \end{figure}
Fig. \ref{fig_ESR_vs_SNR_variation_in_K_and_L_MD_ME_4} depicts the ESR performance versus $\lambda_D$ by changing the number of transmitters and destinations for the OS and SS schemes. The curve for $K=L=1$ serves as a reference to show the performance improvement with selection. We observe that the high-SNR approximate ESR is very close to the exact ESR. We also find that the increase in the number of  transmitters or destinations improves the ESR performance due to the diversity benefit of the selection. As expected, the OS scheme has the best performance among the two schemes. It may be noted that the ESR curve corresponding to $\{K, L\}=\{3, 1\}$ in the SS scheme also corresponds to $\{K, L\}=\{1, 3\}$ in the SS scheme as well as the OS scheme. It is because $K=1$ corresponds to the destination selection only. 
% It is noted that the ESR performance for $\{K, L\}=\{1, 3\}$ is the same for the OS and SS schemes as $K=1$ means destination selection only.  
We also notice that with an increase in $K$ from 1 to 3 for any $L=\{1, 3\}$, the ESR performance improvement in the OS scheme is more significant than in the SS scheme. Furthermore, we observe that with an increase in $L$ from $1$ to $3$, although the OS scheme outperforms the SS scheme for any $K=\{1, 3\}$, this advantage is not as significant as in the case of increasing $K$ for a fixed value of $L$. These observations show that the OS scheme can take advantage of increasing both $K$ and $L$; however, increasing $K$ is better for system security than increasing $L$ in the OS scheme.

For the SS scheme, we notice that the ESR curves for $\{K,L\}=\{1,3\}$ and $\{K,L\}=\{3,1\}$ overlap. This is because in both cases there are $K\times L=3$ available i.i.d. links from which to select the best source-destination pair. This also suggests that if the total number of links $KL$ is the same for any combination of $K$ and $L$, the ESR performance is the same for these combinations in the SS scheme. However, in contrast to the SS scheme, the curves for these combinations in the OS scheme do not overlap. The reason is that the selection also considers the eavesdropper channel quality and hence does not provide the same source-destination pair at all times for these combinations. It is also noted that the SS scheme can outperform the OS scheme by increasing the number of destinations only, thereby reducing the reliance on the eavesdropping channel feedback in the OS scheme, which may not be available for selection.

Fig. \ref{ESR_vs_SNR_SS_OS_High_SNR_Asymp_variation_in_K_L_and_both_M} plots the high-SNR approximate ESR and its asymptotes with respect to $\lambda_D$ for $K=L \in\{1,2\}$ and $M_D=M_E\in\{1,2\}$. As the high-SNR approximate ESR is very close to the exact ESR, only the approximate ESR curves are included in this figure for clarity of presentation. 
% Also, by setting  a high value of $K=9$, we ensure better clarity without disturbing fundamental observations.
% are able to compare the performance gap between the OS and SS  for a large number of transmitters in a better way.
The asymptotes closely match the ESR at high SNR, which validates our theoretical analysis. 
We observe that all the curves have the same slope of unity irrespective of the values of $\lambda_E$, $K$, $L$, $M_D$, and $M_E$.
% , as we derived in Section \ref{subsection_Asymptotic_Analysis_of_ESR_in_OS} and \ref{subsection_Asymptotic_Analysis_of_ESR_in_SS}.  
As the number of multipath components is increased while maintaining $M_D=M_E$, 
% to $M_D=M_E=2$, 
we find that the performance degrades in both selection schemes for given values of $K$ and $L$. Also, we note that the performance in the case of $M_D=M_E=1$ (i.e., for the Rayleigh fading condition) is the best for given values of $K$ and $L$.
For the case $M_D=M_E$, as the number of multipath components in both the destination and eavesdropper channel increases, the ESR performance always degrades.
% However, this is not obvious, as we expect that the performance should improve as $M_D$ increases for a given $M_E$ and the performance should degrade as $M_E$ increases for a given $M_D$. 
This shows that the performance of the selection schemes is dominated by $M_E$, or equivalently by the eavesdropping channel quality. The system with the worst eavesdropper channel (i.e., a Rayleigh fading channel to the eavesdropper) has the best ESR performance.

\begin{figure}
\centering
     \includegraphics[width=3in]{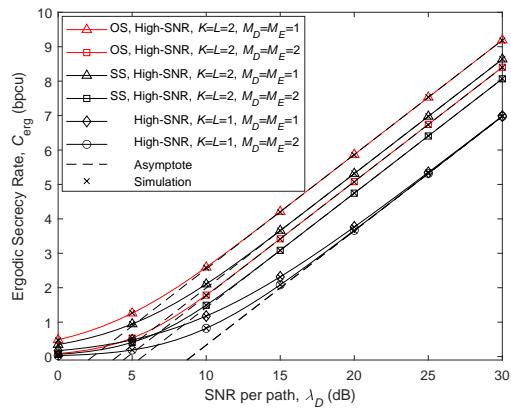} 
 \caption{Variation of the high-SNR approximate ESR and its asymptotic performance with $\lambda_D$ for $\lambda_E=9$ dB.}
 %\vspace{-0.3cm}
 \label{ESR_vs_SNR_SS_OS_High_SNR_Asymp_variation_in_K_L_and_both_M}
 \end{figure}

Furthermore, as $\lambda_D$ improves, the ESR curves corresponding to different numbers of multipath components merge at high SNR when $K=L=1$. This is because the high-SNR offset for the asymptotic ESR in the case of $K=L=1$ for any $M_D=M_E$ is the same. This is not the case when $K>1$ and $L>1$; for example, it can be seen that for $K=L=2$, the curves for $M_D=M_E=1$ and $M_D=M_E=2$ do not merge. These two observations corroborate our analysis regarding the high-SNR offset for any $K$ and $L=1$ in the OS scheme and any $K$ and $L$ in the SS scheme provided in Section \ref{subsection_Asymptotic_Analysis_of_ESR_in_OS} and  \ref{subsection_Asymptotic_Analysis_of_ESR_in_SS}, respectively. 
% \textcolor{red}{ and $K,L>1$ (with $M_D=M_E$) }

 \begin{figure}
 \centering
    \includegraphics[width=3.2in]{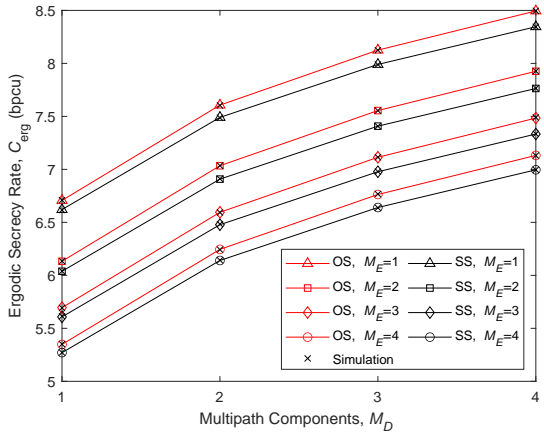} 
 \caption{Variation of exact ESR  with $M_D$ for $K=L=2$, $\lambda_E=0$ dB, and $\lambda_D=20$ dB.}
\vspace{-0.5cm}
 \label{fig_ESR_vs_M_D_SS_OS_fixed_K=L=2}
 \end{figure}
To understand the effect of the number of multipath components on the ESR, we show the variation of the ESR with $M_{D}$ for different values of $M_E$ at $K=L=2$, $\lambda_E=0$ dB and $\lambda_D=20$ dB in Fig. \ref{fig_ESR_vs_M_D_SS_OS_fixed_K=L=2}. For a given set of parameters, as we increase $M_D$, the ESR increases, whereas an increase in $M_E$ causes a decrease in the ESR. 
This occurs because an increase in $M_D$ and $M_E$ increases the received SNR at $D$ and $E$, respectively, due to the improvement in the corresponding channel quality.
% We can also observe that as $M_D$ increases, the ESR increases for both the OS and SS scheme. This is reasonable as main channel received SNR improves as channel quality improves. 
Further, the performance gap between the OS and SS schemes also improves with increasing $M_D$, though very slowly.
The performance gap between the OS and SS schemes also increases with a decrease in $M_E$. This suggests that the OS scheme can take better advantage of the eavesdropping quality degradation. As $M_E=1$ corresponds to a Rayleigh faded eavesdropper channel, this condition represents the worst possible channel for an eavesdropper. These observations suggest that the OS scheme is more sensitive to the variation in $M_E$ than the variation in $M_D$.

Fig. \ref{CORR_graph} shows the ESR performance versus $\lambda_D$ for different levels of transmitter correlation and path correlation for the OS and SS schemes. Figs. \ref{fig_ESR_Corr_vs_SNR_OS_K_L_MD_ME_4_rho_E_fixed_0} and  \ref{fig_ESR_Corr_vs_SNR_OS_K_L_MD_ME_4_rho_D_fixed_0} are for the OS scheme, while Figs. \ref{fig_ESR_Corr_vs_SNR_SS_K_L_MD_ME_4_rho_E_fixed_0} and  \ref{fig_ESR_Corr_vs_SNR_SS_K_L_MD_ME_4_rho_D_fixed_0} represent the SS scheme. Figs. \ref{fig_ESR_Corr_vs_SNR_OS_K_L_MD_ME_4_rho_E_fixed_0} and  \ref{fig_ESR_Corr_vs_SNR_SS_K_L_MD_ME_4_rho_E_fixed_0} show 
% transmitter correlation and the correlation between paths towards destinations by 
variations in $\rho_S$ and $\rho_D$ 
% with no correlation between paths towards eavesdropper i.e.,  
while keeping $\rho_E=0$. In Figs. \ref{fig_ESR_Corr_vs_SNR_OS_K_L_MD_ME_4_rho_D_fixed_0} and \ref{fig_ESR_Corr_vs_SNR_SS_K_L_MD_ME_4_rho_D_fixed_0}, $\rho_S$ and $\rho_E$ are varied while keeping $\rho_D=0$. 

In all of these figures, it can be seen that as the transmitter correlation coefficient is increased from $\rho_S=0$ to $\rho_S=0.9$, the ESR performance degrades. The effect of transmitter correlation in the OS scheme is greater 
 as observed in Figs. \ref{fig_ESR_Corr_vs_SNR_OS_K_L_MD_ME_4_rho_E_fixed_0} and \ref{fig_ESR_Corr_vs_SNR_OS_K_L_MD_ME_4_rho_D_fixed_0} 
 as compared to the SS scheme in Figs.  \ref{fig_ESR_Corr_vs_SNR_SS_K_L_MD_ME_4_rho_E_fixed_0} and \ref{fig_ESR_Corr_vs_SNR_SS_K_L_MD_ME_4_rho_D_fixed_0}. 

\begin{figure*}[!t]
\centering
\subfloat[]{\includegraphics[width=2.8in]{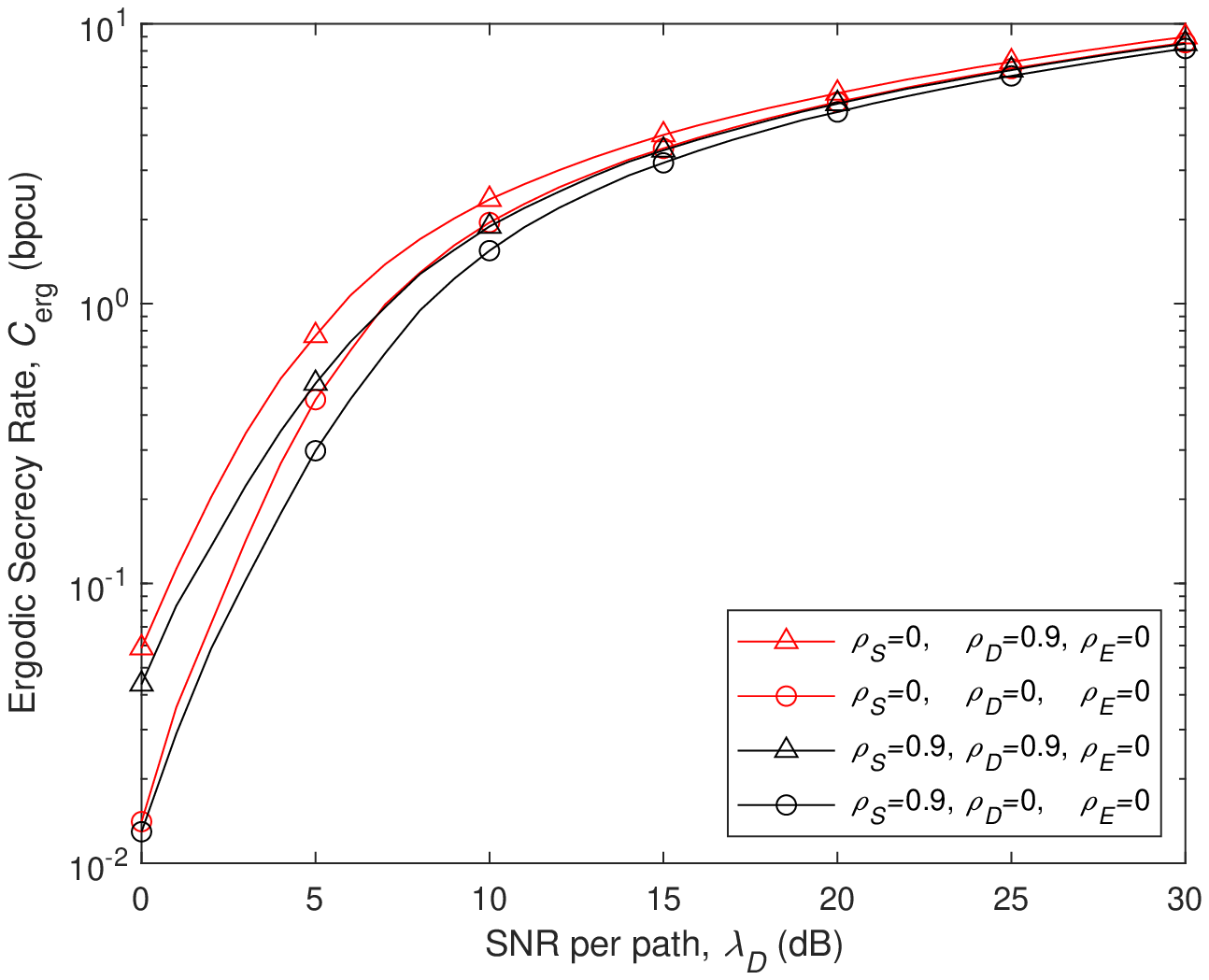}%
\label{fig_ESR_Corr_vs_SNR_OS_K_L_MD_ME_4_rho_E_fixed_0}}
\hfil
\subfloat[]{\includegraphics[width=2.8in]{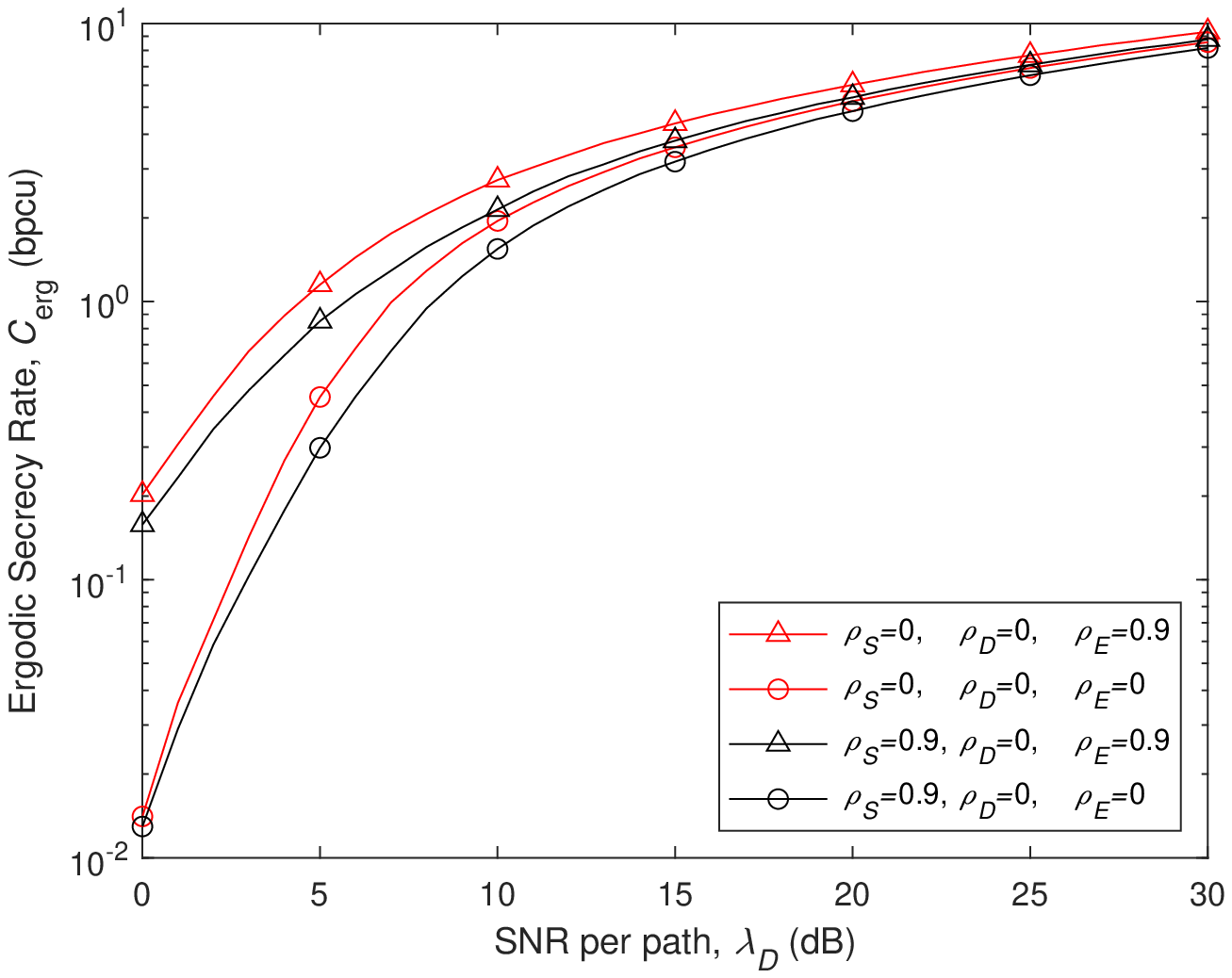}
\label{fig_ESR_Corr_vs_SNR_OS_K_L_MD_ME_4_rho_D_fixed_0}}
\hfil
\subfloat[]{\includegraphics[width=2.8in]{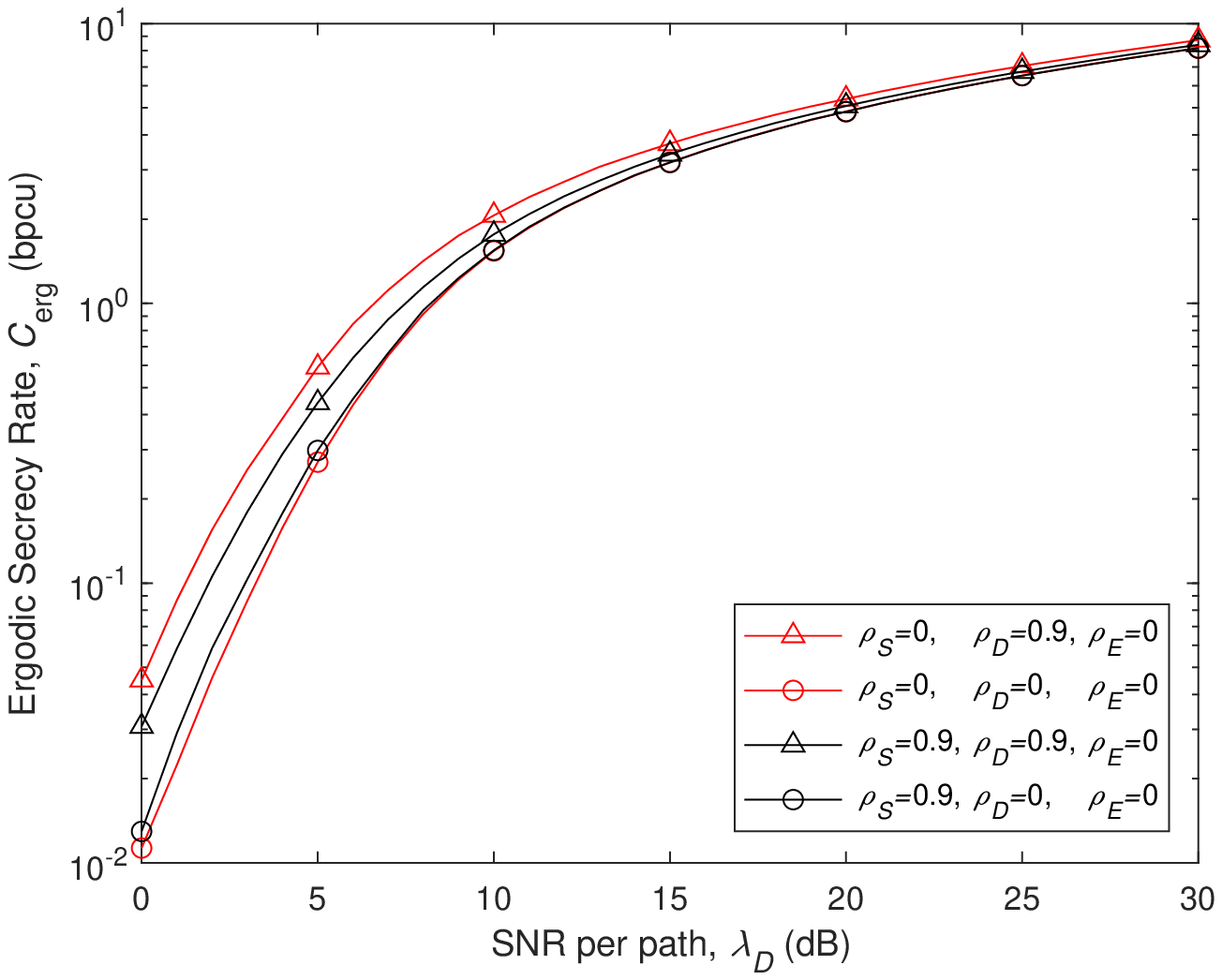}
\label{fig_ESR_Corr_vs_SNR_SS_K_L_MD_ME_4_rho_E_fixed_0}}
\hfil
\subfloat[]{\includegraphics[width=2.8in]{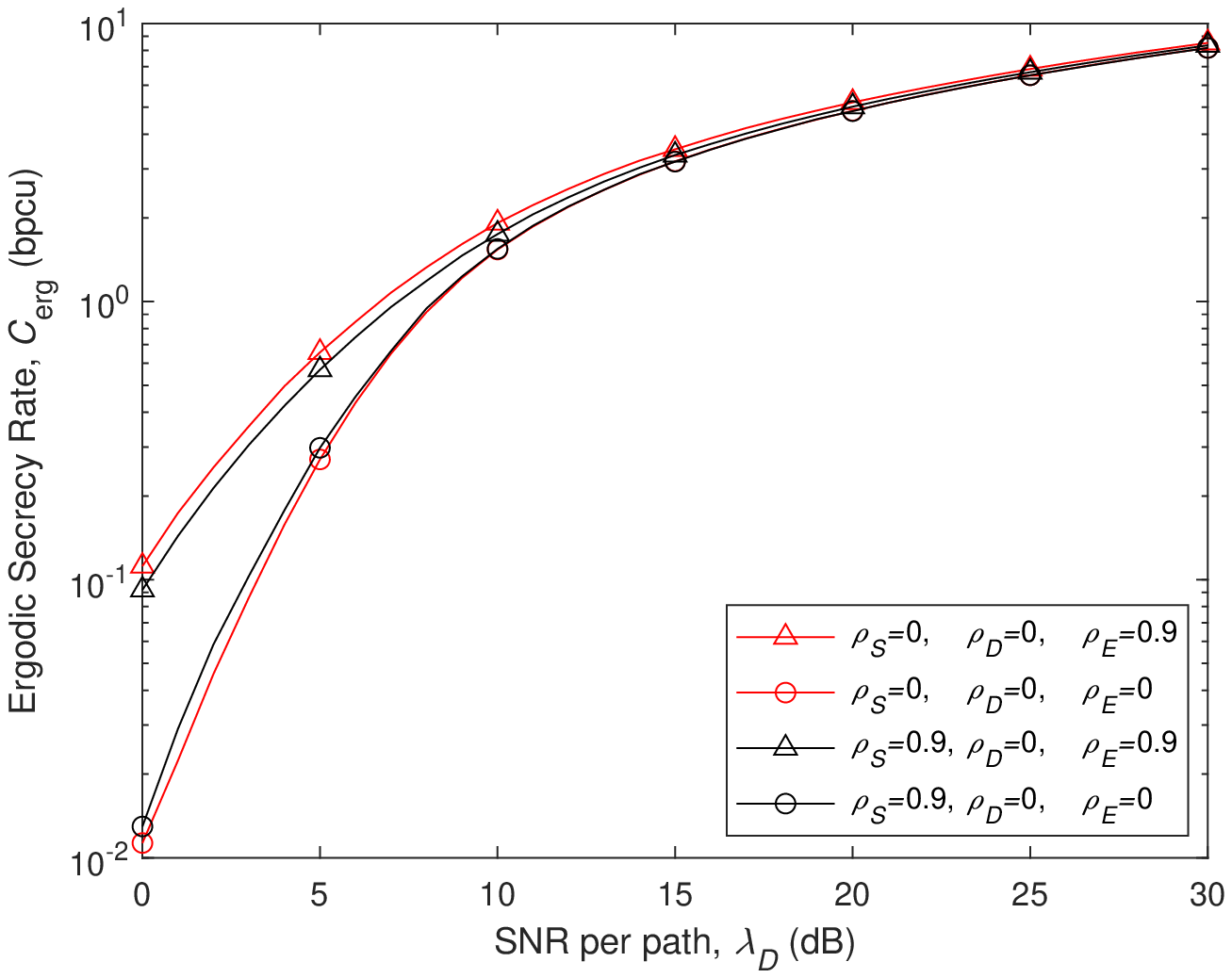}
\label{fig_ESR_Corr_vs_SNR_SS_K_L_MD_ME_4_rho_D_fixed_0}}
\caption{Variation of the exact ESR with $\lambda_D$ for $K=L=4$, $M_D=M_E=4$, and $\lambda_E=9$ dB. (a) OS, fixed $\rho_E$. (b) OS, fixed $\rho_D$. (c) SS, fixed $\rho_E$. (d) SS, fixed $\rho_D$.}
\label{CORR_graph}
\end{figure*}
From Fig. \ref{fig_ESR_Corr_vs_SNR_OS_K_L_MD_ME_4_rho_E_fixed_0} and  \ref{fig_ESR_Corr_vs_SNR_SS_K_L_MD_ME_4_rho_E_fixed_0}, we notice that increasing $\rho_D$ improves the ESR performance in both the OS and SS schemes. This is a result of power focusing on the destination due to the reduced effective dimensionality when the correlation increases. Effective dimensionality reduction was also observed in \cite{PLS_TAS_Corr_2013} in the case of increased destination antenna correlation (this was for the case of a Nakagami flat fading channel). The performance improvement reduces with increasing $\lambda_{D}$ as all of the curves merge at high SNR. This indicates that at high SNR, effective dimensionality reduction is insignificant. 

We also observe that an increase in $\rho_E$ (in Fig. \ref{fig_ESR_Corr_vs_SNR_OS_K_L_MD_ME_4_rho_D_fixed_0} for the OS scheme and in Fig. \ref{fig_ESR_Corr_vs_SNR_SS_K_L_MD_ME_4_rho_D_fixed_0} for the SS scheme) improves the secrecy performance.
% This observation is also similar to that made in \cite{PLS_TAS_Corr_2013} where the eavesdropper antenna correlation is increased. 
Although an increase in both $\rho_D$ and $\rho_E$ improves secrecy performance, by comparing Fig. \ref{fig_ESR_Corr_vs_SNR_OS_K_L_MD_ME_4_rho_E_fixed_0} and Fig. \ref{fig_ESR_Corr_vs_SNR_SS_K_L_MD_ME_4_rho_E_fixed_0} 
 with Fig. \ref{fig_ESR_Corr_vs_SNR_OS_K_L_MD_ME_4_rho_D_fixed_0} and Fig. \ref{fig_ESR_Corr_vs_SNR_SS_K_L_MD_ME_4_rho_D_fixed_0}, we notice that an increase in $\rho_E$ is better for secrecy than an increase in $\rho_D$ in both the OS and SS schemes, irrespective of the value of $\rho_S$. 

\color{black}

\section{Conclusion}\label{sec_Conclusion}
In this paper, the ESR of optimal and sub-optimal source-destination pair selection schemes is evaluated in closed form  for a system with multiple transmitters, multiple destinations, and a single eavesdropper in frequency selective fading channels with SC-CP signaling. We also numerically demonstrate the effect of transmitter correlation along with destination and eavesdropper path correlation. A high-SNR analysis and an asymptotic analysis have been provided to yield a low-complexity ESR evaluation and provide key insights. Our analysis also provides the corresponding ESR results in narrowband Nakagami fading channels with arbitrary integer parameter $m$. Our proposed analysis is general and can also be used to evaluate the ESR of other sub-optimal selection schemes already available in the literature. We notice that the highest ESR is achieved under Rayleigh fading conditions irrespective of the selection schemes for any given combination of the number of transmitters and destinations. 
We also find that the OS scheme can take better advantage of the multipath conditions in the system.
We observe that for the OS scheme, it is better to increase the number of transmitters than destinations for the improvement of the ESR, while on the other hand, in the case of the SS scheme, the ESR shows the same dependence on the number of transmitters as on the number of destinations. We also observe that while transmitter correlation deteriorates the ESR, destination and eavesdropper path correlation both improve it.  Future work includes studying the impact of imperfect/outdated CSI on the SOP and ESR performance of source-destination selection in frequency selective fading channels, as well as extending the analysis to the case where multiple eavesdroppers are present in the system.

% \section{Appendix}
\appendix
\subsection{Evaluation of $J_0^{(k)}$  in (\ref{eq_J_0_OS_start_eqn}).}
\label{appendix1}
In the denominator of (\ref{eq_J_0_OS_start_eqn}), the values of $l_q$ for different $q\in\{1 ,\ldots,  k\}$ may be the same or different. We denote by  $\mathcal{I}$ the number of distinct values of $l_q$ that are taken on more than once. We also define $\mathcal{Q}_i$, where $i\in\{1 ,\ldots, \mathcal{I}\}$, as the set of indices $q$ for which $l_q$ takes the $i^{\text{th}}$ distinct value. We also define $\mathcal{Q}=\mathcal{Q}_1\cup\mathcal{Q}_2\cup ,\ldots, \mathcal{Q}_\mathcal{I}$ and $\bar{\mathcal{Q}}=\{1 ,\ldots, k\}-\mathcal{Q}$.

Note that when the values of $l_q$ for each $q\in\{1,\ldots,k\}$ are distinct, $\bar{\mathcal{Q}}$ is an empty set. Using the above notations, the partial fraction expansion of the integrand in (\ref{eq_J_0_OS_start_eqn}) is obtained, and then the solution is obtained with the help of the integral solution \cite[eq. (3.462.16) and (3.462.19)]{ryzhik_2007} as
% \begin{table*}
% \begin{strip}
\begin{align}
\label{eq_J_0_OS_solution}
J_0^{(k)}&=\int_1^\infty\Big(\frac{A\exp({-\frac{\widetilde{l}^{(k)}}{\lambda_{D}}x})}{x}\nn\\
&+\sum_{i=1}^{\mathcal{I}}\sum_{t=1|q\in\mathcal{Q}_i}^{|\mathcal{Q}_i|M_E+\sum_{q\in\mathcal{Q}_i}\widehat{n}_{q}^{(l_q)}}\frac{B_{i,t}\exp({-\frac{\widetilde{l}^{(k)}}{\lambda_{D}}x})}{\big(x+\frac{\lambda_{D}}{l_q\lambda_{E}}\big)^{t}} \nn\\
&+\sum_{q\in\bar{\mathcal{Q}}}\sum_{t=1}^{M_E+\widehat{n}_{q}^{(l_q)}}\frac{C_{q,t}\exp({-\frac{\widetilde{l}^{(k)}}{\lambda_{D}}x})}{\big(x+\frac{\lambda_{D}}{l_q\lambda_{E}}\big)^{t}}\Big)dx.
\end{align}
% \hrule
% \end{table*}
The coefficients $A$, $B_{i,t}$, and $C_{q,t}$ are evaluated using the method of partial fractions. The second term in (\ref{eq_J_0_OS_solution}) corresponds to the case of values $l_q$ that appear more than once, while the third term corresponds to the case of values of $l_q$ that appear exactly once. For given values of $K$, $L$, $M_D$, $M_E$, the partial fraction coefficients $A$, $B_{i,t}$ and $C_{q,t}$ are easily obtained.

\subsection{Evaluation of  $J_1^{(k)}$ in  (\ref{eq_J_1_OS_start_eqn}).}
\label{appendix2}
The solution of $J_1^{(k)}$ in (\ref{eq_J_1_OS_start_eqn}) also depends on how many $l_q$ for different $q\in\{1 ,\ldots, k\}$ in the denominator of (\ref{eq_J_1_OS_start_eqn}) are equal, as in  (\ref{eq_J_0_OS_start_eqn}) for $J_0^{(k)}$. In addition,  the solution method also needs to take care that the power of $x$ in the numerator of (\ref{eq_J_1_OS_start_eqn}) may be greater than the highest power of $x$ in the denominator; note that this was not an issue for $J_0^{(k)}$ in (\ref{eq_J_0_OS_start_eqn}). The solution method is first to check whether all $l_q$ are equal to each other in (\ref{eq_J_1_OS_start_eqn});  if they are all equal, no partial fraction is required. We then adopt a change of variable, considering that the power of $x$ in the numerator can be greater than that in the denominator. Finally, using the integration solution \cite[eq. (3.351.2)]{ryzhik_2007} we obtain the solution as

\begin{align}
% \label{eq_I_0_OS_start_eqn}
% I_0^{(k)}&=\int_1^\infty\frac{\exp({-\frac{\widetilde{l}^{(k)}}{\lambda_{D}}x})}{x\Big(\prod\limits_{q=1}^{k}\big(x+\frac{\lambda_{D}}{l_q\lambda_{E}}\big)^{M_E+\widehat{n}_{q}^{(l_q)}}\Big)}dx\\ 
\label{eq_J_1_OS_start_eqn_appendix_1}
% J_1^{(k)}&=\int_1^\infty\frac{x^{\widetilde{m}^{(k)}-\widetilde{u}^{(k)}-1}\exp({-\frac{\widetilde{l}^{(k)}}{\lambda_{D}}x})}{\prod_{q=1}^{k}\big(x+\frac{\lambda_{D}}{l_q\lambda_{E}}\big)^{M_E+\widehat{n}_{q}^{(l_q)}}}dx\nn\\
% J_1^{(k)}&=\int_1^\infty\frac{x^{\widetilde{m}^{(k)}-\widetilde{u}^{(k)}-1}\exp({-\frac{\widetilde{l}^{(k)}}{\lambda_{D}}x})}{\big(x+\frac{\lambda_{D}}{l_q\lambda_{E}}\big)^{kM_E+\sum_{q=1}^{k}\widehat{n}_{q}^{(l_q)}}}dx\nn\\
J_1^{(k)}&=\sum_{j=0}^{\widetilde{m}^{k}-\widetilde{u}^{k}-1}\binom{\widetilde{m}^{k}-\widetilde{u}^{k}-1}{j}  \nn\\
& \times \Big(-\frac{\lambda_{D}}{l_q\lambda_{E}}\Big)^{\widetilde{m}^{k}-\widetilde{u}^{k}-1-j}\exp\Big({\frac{\widetilde{l}^{(k)}}{l_q\lambda_{E}}}\Big)
\nn\\
&\times 
\int_{1+\frac{\lambda_{D}}{l_q\lambda_{E}}}^\infty z^{j-(kM_E+\sum_{q=1}^{k}\widehat{n}_{q}^{(l_q)})}\exp\Big({-\frac{\widetilde{l}^{(k)}}{\lambda_{D}}z}\Big)dz.
\end{align}

In the case where the $l_q$ values for distinct $q\in\{1 ,\ldots, k\}$ are not all equal in
(\ref{eq_J_1_OS_start_eqn}), we use partial fractions without including the numerator. The partial fraction method is similar to that used in (\ref{eq_J_0_OS_solution}). Then we adopt a change of variable similar to (\ref{eq_J_1_OS_start_eqn_appendix_1}) and subsequently use the  integration solution \cite[eq. (3.351.2)]{ryzhik_2007} to obtain the solution as 
\begin{align}
\label{eq_J_1_OS_start_eqn_appendix_2}
J_1^{(k)}
% &=\int_1^\infty\frac{x^{\widetilde{m}^{(k)}-\widetilde{u}^{(k)}-1}\exp({-\frac{\widetilde{l}^{(k)}}{\lambda_{D}}x})}{\prod_{q=1}^{k}\big(x+\frac{\lambda_{D}}{l_q\lambda_{E}}\big)^{M_E+\widehat{n}_{q}^{(l_q)}}}dx\nn\\
&=\sum_{i=1}^{\mathcal{I}}\sum_{t=1|q\in\mathcal{Q}_i}^{|\mathcal{Q}_i|M_E+\sum_{q\in\mathcal{Q}_i}\widehat{n}_{q}^{(l_q)}}\sum_{j=0}^{\widetilde{m}^{k}-\widetilde{u}^{k}-1}\exp\Big({\frac{\widetilde{l}^{(k)}}{l_q\lambda_{E}}}\Big)\nn\\
& \times \binom{\widetilde{m}^{k}-\widetilde{u}^{k}-1}{j}\Big(-\frac{\lambda_{D}}{l_q\lambda_{E}}\Big)^{\widetilde{m}^{k}-\widetilde{u}^{k}-1-j} B_{i,t}\nn\\
 &\times\int_{1+\frac{\lambda_{D}}{l_q\lambda_{E}}}^\infty z^{j-t}\exp\Big({-\frac{\widetilde{l}^{(k)}}{\lambda_{D}}z}\Big)dz \nn\\  &+\sum_{q\in\bar{\mathcal{Q}}}\sum_{t=1}^{M_E+\widehat{n}_{q}^{(l_q)}}\sum_{j=0}^{\widetilde{m}^{k}-\widetilde{u}^{k}-1}\exp\Big({\frac{\widetilde{l}^{(k)}}{l_q\lambda_{E}}}\Big)\nn\\
 &\times \binom{\widetilde{m}^{k}-\widetilde{u}^{k}-1}{j}\Big(-\frac{\lambda_{D}}{l_q\lambda_{E}}\Big)^{\widetilde{m}^{k}-\widetilde{u}^{k}-1-j} C_{q,t}\nn\\
 &\times \int_{1+\frac{\lambda_{D}}{l_q\lambda_{E}}}^\infty z^{j-t}\exp\Big({-\frac{\widetilde{l}^{(k)}}{\lambda_{D}}z}\Big)dz.
\end{align}
Here, $\mathcal{I}$, $\mathcal{Q}_i$, and $\bar{\mathcal{Q}}$ have same definition as in (\ref{eq_J_0_OS_solution}), however, the partial fraction coefficients $B_{i,t}$ and $C_{q,t}$ are specific to $J_1^{(k)}$. These coefficients are evaluated in a manner similar to (\ref{eq_J_0_OS_solution}).

\subsection{Evaluation of $J_1^{(k)}$ in (\ref{high_snr_J1}) at high SNR.}
\label{appendix4}
When all $l_q$ are equal for all $q\in\{1 ,\ldots, k\}$, the solution of $J_1^{(k)}$ in (\ref{high_snr_J1}) at high SNR can be evaluated easily as the numerator's highest power of $x$ is always smaller than any power of $x$ in the denominator. \color{black}The integral is evaluated after a change of variable as
\begin{align}\label{eq_int_J_1_High_SNR_OS_appendix_1}
% J_1^{(k)}&=\int_1^{\infty}\frac{x^{\widetilde{m}^{(k)}-1}}{\prod\limits_{q=1}^{k}\big(x+\frac{\lambda_{D}}{l_q\lambda_{E}}\big)^{M_E+\widehat{m}_{{q}}^{(l_q)}}}dx\nn\\
J_1^{(k)}&=\int_1^{\infty}\frac{x^{\widetilde{m}^{(k)}-1}}{\big(x+\frac{\lambda_{D}}{l_q\lambda_{E}}\big)^{k M_E+\widetilde{m}^{(k)}}}dx.
% \nn\\
% &=\sum_{j=0}^{\widetilde{m}^{(k)}-1}\binom{\widetilde{m}^{(k)}-1}{j}\Big(-\frac{\lambda_{D}}{l_q\lambda_{E}}\Big)^{\widetilde{m}^{(k)}-1-j}\frac{\big(1+\frac{\lambda_D}{l_q\lambda_E}\big)^{-(kM_E+\widetilde{m}^{(k)}-1-j)}}{kM_E+\widetilde{m}^{(k)}-1-j}.
\end{align}
When any $l_q$ is distinct for any $q\in\{1 ,\ldots, k\}$, $J_1^{(k)}$ in (\ref{high_snr_J1}) at high SNR is evaluated by adopting the partial fraction method similar to (\ref{eq_J_0_High_SNR_OS_sol_final}) as  
\begin{align}\label{eq_int_J_1_High_SNR_OS_appendix_2}
% J_1^{(k)}&=\int_1^{\infty}\frac{x^{\widetilde{m}^{(k)}-1}}{\prod\limits_{q=1}^{k}\big(x+\frac{\lambda_{D}}{l_q\lambda_{E}}\big)^{M_E+\widehat{m}_{{q}}^{(l_q)}}}dx\nn\\
J_1^{(k)}&=\int_1^\infty\Big(\sum_{i=1}^{\mathcal{I}}\sum_{t=1|q\in\mathcal{Q}_i}^{|\mathcal{Q}_i| M_E+\sum_{q\in\mathcal{Q}_i}\widehat{m}_{q}^{l_q}}\frac{B_{i,t}}{\big(x+\frac{\lambda_{D}}{l_q\lambda_{E}}\big)^{t}}\nn\\
&+\sum_{q\in\bar{\mathcal{Q}}}\sum_{t=1}^{M_E+\widehat{m}_{q}^{l_q}}\frac{C_{q,t}}{\big(x+\frac{\lambda_{D}}{l_q\lambda_{E}}\big)^{t}}\Big)dx.
\end{align}
The partial fraction coefficients  $B_{i,t}$ and $C_{q,t}$ are obtained using the standard partial fraction technique. 

\bibliographystyle{IEEEtran}
\bibliography{IEEEabrv, ref}
\end{document}